\documentclass[
reprint,
 amsmath,amssymb,
 aps,
floatfix,
]{revtex4-1}

\usepackage{graphicx}
\usepackage{dcolumn}
\usepackage{bm}
\usepackage{physics}

\begin{document}

\preprint{APS/123-QED}

\title{Demonstrating a Continuous Set of Two-qubit Gates for Near-term Quantum Algorithms}

\author{B. Foxen$^*$,$^{1, 2}$ C. Neill$^*$,$^{2}$ A. Dunsworth$^*$,$^{2}$ P. Roushan,$^{2}$ B. Chiaro,$^{1}$ A. Megrant,$^{2}$ J. Kelly,$^{2}$ Zijun Chen,$^{2}$ K. Satzinger,$^{2}$ R. Barends,$^{2}$ F. Arute,$^{2}$ K. Arya,$^{2}$ R. Babbush,$^{2}$ D. Bacon,$^{2}$ J.C. Bardin,$^{2,3}$ S. Boixo,$^{2}$ D. Buell,$^{2}$ B. Burkett,$^{2}$ Yu Chen,$^{2}$ R. Collins,$^{2}$ E. Farhi,$^{2}$ A. Fowler,$^{2}$ C. Gidney,$^{2}$ M. Giustina,$^{2}$ R. Graff,$^{2}$ M. Harrigan,$^{2}$ T. Huang,$^{2}$ S.V. Isakov,$^{2}$ E. Jeffrey,$^{2}$ Z. Jiang,$^{2}$ D. Kafri,$^{2}$ K. Kechedzhi,$^{2}$ P. Klimov,$^{2}$ A. Korotkov,$^{2}$ F. Kostritsa,$^{2}$ D. Landhuis,$^{2}$ E. Lucero,$^{2}$ J. McClean,$^{2}$ M. McEwen,$^{1}$ X. Mi,$^{2}$ M. Mohseni,$^{2}$ J.Y. Mutus,$^{2}$ O. Naaman,$^{2}$ M. Neeley,$^{2}$ M. Niu,$^{2}$ A. Petukhov,$^{2}$ C. Quintana,$^{2}$ N. Rubin,$^{2}$ D. Sank,$^{2}$ V. Smelyanskiy,$^{2}$ A. Vainsencher,$^{2}$ T.C. White,$^{2}$ Z. Yao,$^{2}$ P. Yeh,$^{2}$, A. Zalcman,$^{2}$ H. Neven,$^{2}$ and John M. Martinis$^{1,2,b}$}

\affiliation{%
 $^1$ Department of Physics, University of California, Santa Barbara, CA \\
 $^2$ Google Research, Santa Barbara, CA 93117 \\ 
 $^3$ Department of Electrical and Computer Engineering, University of Massachusetts Amherst, Amherst, MA \\
 $^b$ martinis@physics.ucsb.edu \\
 $^*$ these authors contributed equally to this work.
}%

\collaboration{Google AI Quantum}

\date{\today}

\begin{abstract}
Quantum algorithms offer a dramatic speedup for computational problems in machine learning, material science, and chemistry. However, any near-term realizations of these algorithms will need to be optimized to fit within the finite resources offered by existing noisy hardware.  Here, taking advantage of the adjustable coupling of gmon qubits, we demonstrate a continuous two-qubit gate set that can provide a 3x reduction in circuit depth as compared to a standard decomposition. We implement two gate families: an iSWAP-like gate to attain an arbitrary swap angle, $\theta$, and a CPHASE gate that generates an arbitrary conditional phase, $\phi$. Using one of each of these gates, we can perform an arbitrary two-qubit gate within the excitation-preserving subspace allowing for a complete implementation of the so-called Fermionic Simulation, or fSim, gate set. We benchmark the fidelity of the iSWAP-like and CPHASE gate families as well as 525 other fSim gates spread evenly across the entire fSim($\theta$, $\phi$)  parameter space achieving purity-limited average two-qubit Pauli error of $3.8 \times 10^{-3}$ per fSim gate.


\end{abstract}

\maketitle


\section{\label{sec:Intro}Introduction}
Quantum computing is a potentially transformative technology, but challenges remain in identifying a path towards solving practical problems with a quantum advantage \cite{Feynman1982}. Continued progress towards this goal may be made on many fronts including qubit coherence or scalability \cite{Koch2007,Rosenberg2017}, measurement or gate fidelities \cite{Schuster2005, Jeffrey2014}, and algorithmic improvements that reduce the required circuit depth through compilation \cite{Chong2017}. In superconducting qubits, single-qubit gates are usually a factor of two or more lower error than two-qubit gates. Consequently, a typical strategy has been to demonstrate a minimally universal gate set consisting of arbitrary single-qubit rotations and a single two-qubit gate \cite{PhysRevA.52.3457}. This is an efficient approach for some algorithms, e.g. surface code error correction, which compiles optimally with such a gate set \cite{Fowler2012}. However, many noisy intermediate-scale quantum (NISQ, \cite{Preskill2018}) algorithms require a more diverse set of two-qubit gates. An implementation of these gates could take the place of six to eight single-qubit gates and three $\rm CZ_{\phi}$ gates per arbitrary two-qubit gate required with an optimal decomposition into a minimally-universal gate set \cite{khaneja2001}.

In the NISQ era, we need the largest two-qubit gate set that may be implemented with high-fidelity. A general two-qubit unitary gate allows independent control over the strength of $\sigma_X \sigma_X$, $\sigma_Y \sigma_Y$, and $\sigma_Z \sigma_Z$ coupling between qubits requiring both DC and microwave control of gmon qubits \cite{yuchen2014}. However, models of interacting particles typically conserve the number of excitations corresponding to a simpler model where the $\sigma_X \sigma_X$ and $\sigma_Y \sigma_Y$ couplings have equal coefficients.  This reduces the number of control parameters from three to two and eliminates the need for microwave control during an algorithm. This set of excitation-conserving gates has been appropriately termed the Fermionic Simulation, or fSim, gate set since it maps electron conservation in a chemistry problem to photon conservation in qubits \cite{Kivlichan2018}. An fSim gate can be defined with two control angles, $\theta$, the $\ket{01} \leftrightarrow \ket{10}$ swap angle, and, $\phi$, the phase of the $\ket{11}$ state with a matrix representation in the $\ket{00}, \ket{01}, \ket{10}, \ket{11}$ basis given by:

\begin{eqnarray}
\label{eq:fsim_main}
\rm fSim(\theta, \phi) = \left(\begin{array}{cccc}
     1 & 0 & 0 & 0  \\
     0 & \cos\,\theta & -i\sin\,\theta & 0 \\
     0 & -i\sin\,\theta  & \cos\,\theta & 0 \\
     0 & 0 & 0 & e^{-i\phi}
\end{array}\right)
\end{eqnarray}

\noindent We use this as both a convenient definition and a useful model for describing general two-qubit gates resulting from arbitrary flux control of gmon qubits. Notably, promising low-depth algorithms using this gate set have been proposed including the quantum approximate optimization algorithm \cite{Farhi2016} and an algorithm for linear-depth circuits simulating the electronic structure of molecules \cite{Kivlichan2018}. Additionally, algorithms performed with just z-rotations and fSim gates enable error mitigation techiques including post selection and zero noise extrapolation \cite{Kandala2019}, further improving this gate set’s prospects on NISQ processors.

Here, we first demonstrate the strong flux tunable coupling between gmon qubits which we use to perform fast two-qubit gates. Then, to describe our calibration and control strategy, we use shallow circuits to illuminate the natural correspondence of the coupled transmon Hamiltonian and the fSim gate set. We use cross-entropy benchmarking (XEB, \cite{Boixo2018}) to characterize two linearly independent and continuous families of entangling gates: the iSWAP-like family corresponding to $\rm fSim(\theta, \phi \propto \theta^{2})$, and the CPHASE family corresponding to $\rm fSim(\theta \approx 0^{\circ}, \phi)$. We then combine these two continuous gate sets to calibrate and benchmark 525 fSim gates spread evenly across the entire ($\theta$, $\phi$) parameter space.

\begin{figure}
\includegraphics[width=80mm]{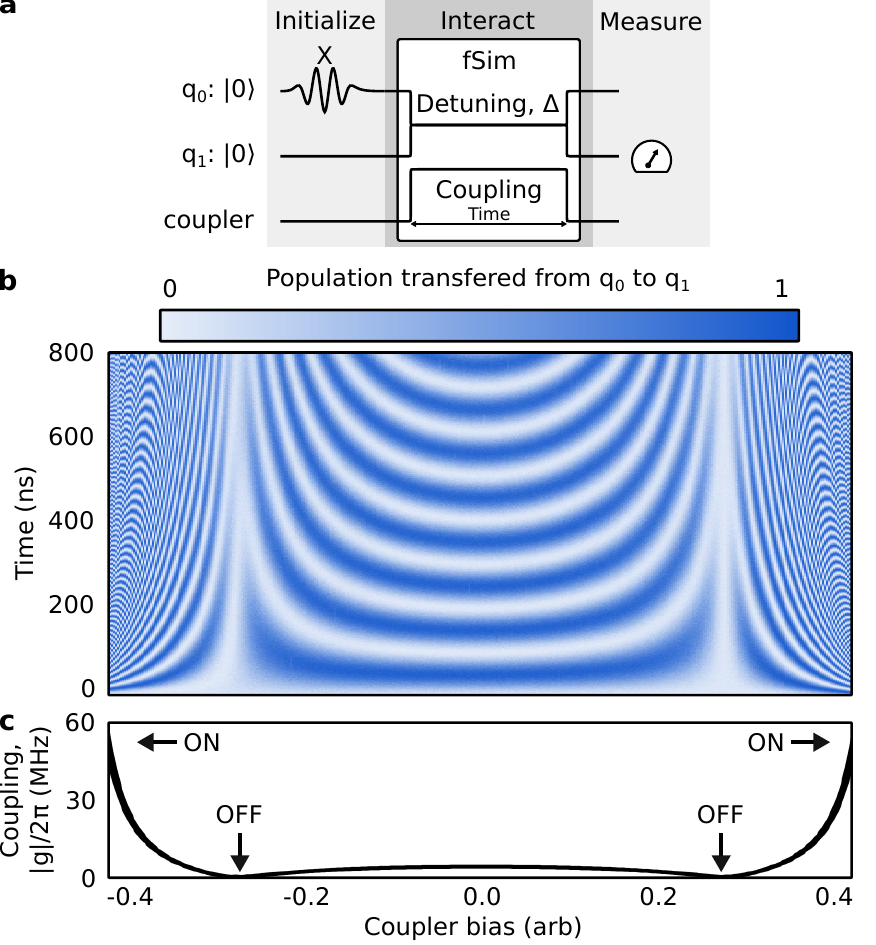}
\caption{\label{fig:coupling}Demonstration of the tunable coupling between gmon qubits. \textbf{a}, Pulse sequence used to measure swapping as a function of coupler bias.  We initialize one qubit, perform an fSim gate, defined by a set of three flux pulses that control the qubit frequencies and the coupling between the qubits, and measure the population of the other qubit. \textbf{b}, We vary the fSim gate as a function of the length and amplitude of the coupler pulse to measure the swap rate as a function of the bias amplitude. \textbf{c}, By taking the Fourier transform of the oscillations in \textbf{b}, we extract the coupling strength, $|g|$, as a function of coupler bias. The coupling changes sign at the two ``OFF'' biases ensuring we can turn the coupling off.  }
\end{figure}

\section{\label{sec:coupling}Strong coupling with gmon qubits}

The quantum processors used in this work each consist of four gmon transmon qubits in a chain, together with three couplers. Both the qubit frequencies and their coupling can be independently controlled, providing several advantages over fixed coupling designs \cite{yuchen2014, charlesthesis, PhysRevApplied.10.054062}. Firstly, since we can turn off the coupling at any detuning, both qubits may idle and perform single-qubit gates while operating closer to their flux insensitive point. This improves dephasing and decreases our sensitivity to flux settling tails. Secondly, since entangling gates are performed by bringing the qubit states near resonance, idling the qubits closer together means that gates require much smaller dynamic detunings, further reducing the amplitude of flux settling tails \cite{rol2019, foxen2018}. Thirdly, since the on/off coupling ratio is not dependent on the maximum qubit-qubit detuning, we are able to increase the overall coupling strength enabling faster gates with reduced decoherence error.

In Figure \ref{fig:coupling} we characterize the qubit-qubit coupling strength as a function of the coupler flux bias. Using the pulse sequence in Figure \ref{fig:coupling}a, we initialize one qubit, apply an fSim gate, and measure the population transferred to the other qubit. Each fSim gate is defined by the amplitude and duration of three, nominally rectangular, flux bias pulses. Two pulses control the qubit frequencies and set their relative detuning, $\Delta$, while the third pulse controls the coupling strength between the qubits, $g$. In Figure \ref{fig:coupling}b we repeat this pulse sequence using the qubit flux biases to place them on resonance ($\rm \Delta = 0\,$MHz) while varying the coupler bias amplitude and the shared duration of all three pulses. By taking  the Fourier transform of the oscillating population data in \ref{fig:coupling}b we extract the swap rate as a function of coupler bias which is equivalent to twice the qubit-qubit coupling, $g$, plotted in Figure \ref{fig:coupling}c. We measure $g/2\pi=6\,$MHz when the coupler is biased to zero $\Phi_{0}$, and a coupling exceeding -50\,MHz as the coupler bias approaches $\rm \Phi_{0}/2$. The net coupling changes sign between these two regions ensuring we can turn the coupling off. During general operation, we idle and perform single-qubit gates with the coupler at the ``OFF'' bias and make excursions to stronger couplings (``ON'' region) to perform fSim gates. In this work we use $g_{\rm{max}}/2\pi \approx -45\,$MHz which is three times stronger than is typical for fixed coupling devices.

\section{\label{sec:transmon_physics}Coupled Transmon physics and the \lowercase{f}S\lowercase{im} gate set}

\begin{figure*}
\includegraphics[width=183mm]{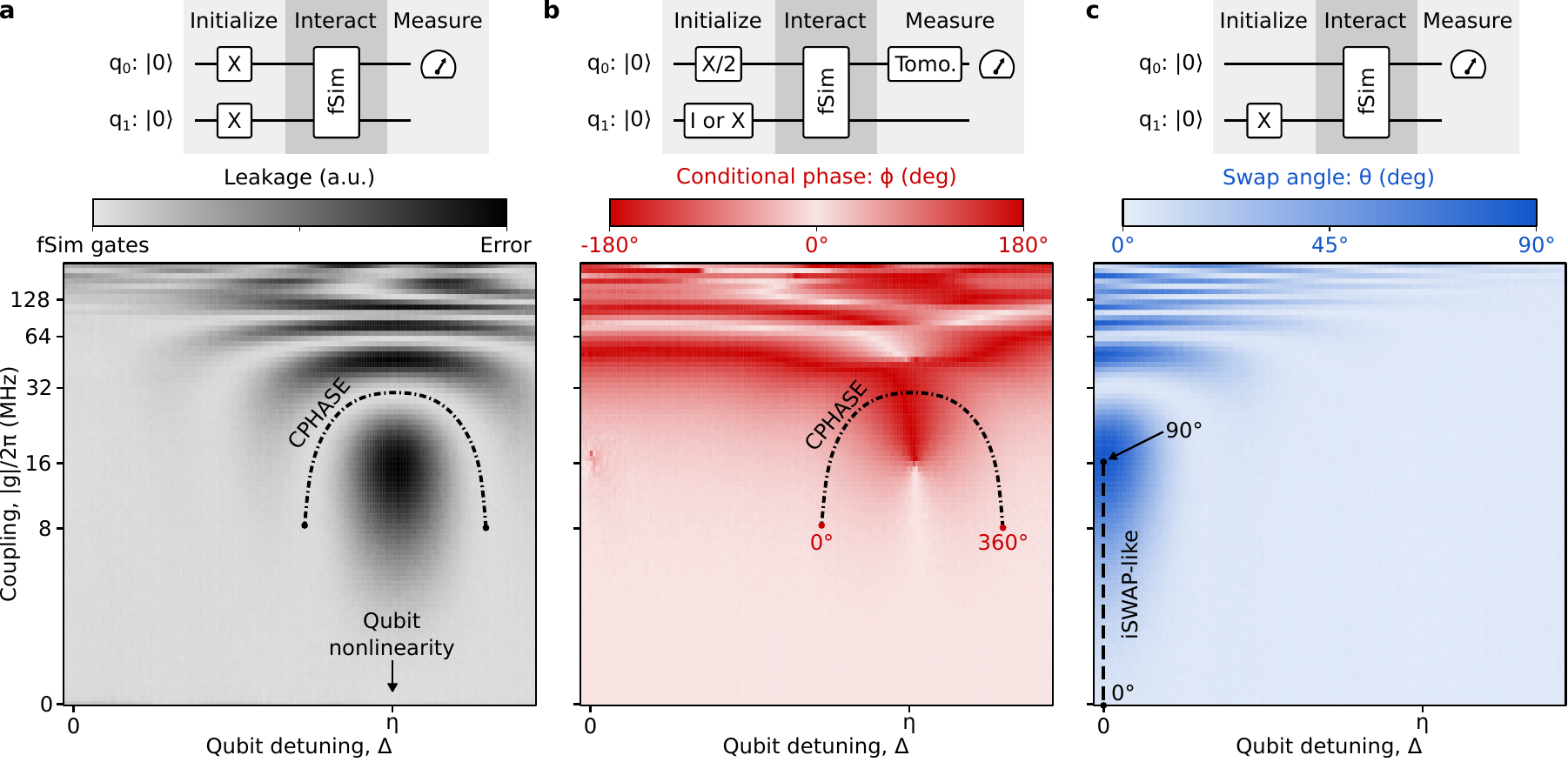} \caption{Exploring the parameter space of two-qubit gates. Each pixel represents one experiment. We use a set of 15\,ns rectangular current bias waveforms to perform some fSim unitary by setting the qubit-qubit detuning, $\Delta$, and the coupling strength, $g$. \textbf{a}, To identify the low-leakage gates described by the fSim model, we measure leakage by initializing the $\ket{11}$ state and measuring the $\ket{02}$ state. When the detuning is near the qubit nonlinearity, we observe the expected Rabi oscillations. \textbf{b}, We measure the conditional phase, $\phi$, by performing a Ramsey experiment where we initialize one qubit with an X/2 gate and perform tomography to measure the difference in accumulated phase ($\phi$) with and without initializing the other qubit to the $\ket{1}$ state. By choosing combinations of $\Delta$ and $g$ as indicated by the CPHASE dash-dotted line (chosen as the low-leakage coupling strength from \textbf{a}), we are able to achieve any $\rm \phi : [-180^{\circ}, 180^{\circ}]$. \textbf{c}, We measure the swap angle, $\theta$, by initializing the $\ket{01}$ state and measuring the $\ket{10}$ state. By placing the qubits on resonance and varying the coupling strength along the iSWAP-like dashed line, we are able to achieve any $\rm \theta : [0^{\circ}, 90^{\circ}]$.
\label{fig:background_scans}}
\end{figure*}

In the absence of a resonant microwave drive, coupled transmon qubits naturally evolve within the excitation-preserving subspace. The specific time evolution is determined by three parameters: the qubit nonlinearity, $\eta$, the qubit-qubit frequency detuning,$\Delta$, and the coupling between qubits, $g$. While $\eta$ is fixed at 240\,MHz by qubit capacitance, the gmon architecture allows for time-dependent control of both $\Delta$ and $|g|$ using DC to $\approx 200\,$MHz bandwidth flux waveforms. The qubit center frequency, $(f_{q1} + f_{q2})/2$, is a free parameter that may be used to avoid coupled two level system (TLS) defects present in the frequency spectrum of either qubit \cite{Macha2010, Shalibo2010, Klimov2018}. For simplicity, we limit our fSim control pulses to synchronous, nominally rectangular waveforms defined by four parameters: a shared length, typically 13\,ns to 15\,ns and three control amplitudes that set $g$ and $\Delta$. While further pulse shaping may improve gate performance in the future, these basic waveforms were sufficient to approach the decoherence limit of our qubits which have a $\rm T_1$ of $25.3 \pm 7.3\,\mu$s (supplement \ref{app:T1}).

The full fSim control model describes any low-leakage two-qubit unitary evolution with five parameters: $\theta$ and $\phi$, discussed previously, in addition to three parameters describing single-qubit phases as detailed in the supplement (\ref{app:fSim_parameterization}). Here, we focus on the parameters that describe the two-qubit interaction and use the three experiments described in Figure \ref{fig:background_scans} to measure leakage to the $\ket{02}$ state and map out the $\phi$ and $\theta$ control landscape (complete unitary tomography procedure outlined in supplemental section \ref{app:unitary_tomography}). Each experiment follows the same pattern: initialize a relevant state, apply fSim control pulses, and then perform either population or tomographic measurements to extract the desired qubit's population or phase. Within the fSim model, leakage is the dominant error. In Figure \ref{fig:background_scans}a, we map out leakage by initializing $\ket{11}$ and measuring the $\ket{0}$ population of the lower frequency qubit as a proxy for leakage in the higher frequency qubit. In Figure \ref{fig:background_scans}b we explore the $\phi$ parameter space by performing a Ramsey experiment where we take the difference in the accumulated phase with and without the second qubit initialized to the $\ket{1}$ state. Finally, in Figure \ref{fig:background_scans}c we explore the $\theta$ parameter space by initializing one qubit to the $\ket{1}$ state and measuring the $\ket{1}$ population of the other qubit after the fSim gate. The Rabi oscillation physics explored with these measurements is reproduced with fairly rudimentary numerics in supplemental \ref{app:numerics}, but these experiments serve to demonstrate our fSim control strategy.

\section{\label{sec:cphase_iswap_like}Benchmarking \lowercase{i}SWAP-\lowercase{like} and CPHASE gates}
\begin{figure}
\includegraphics[width=80mm]{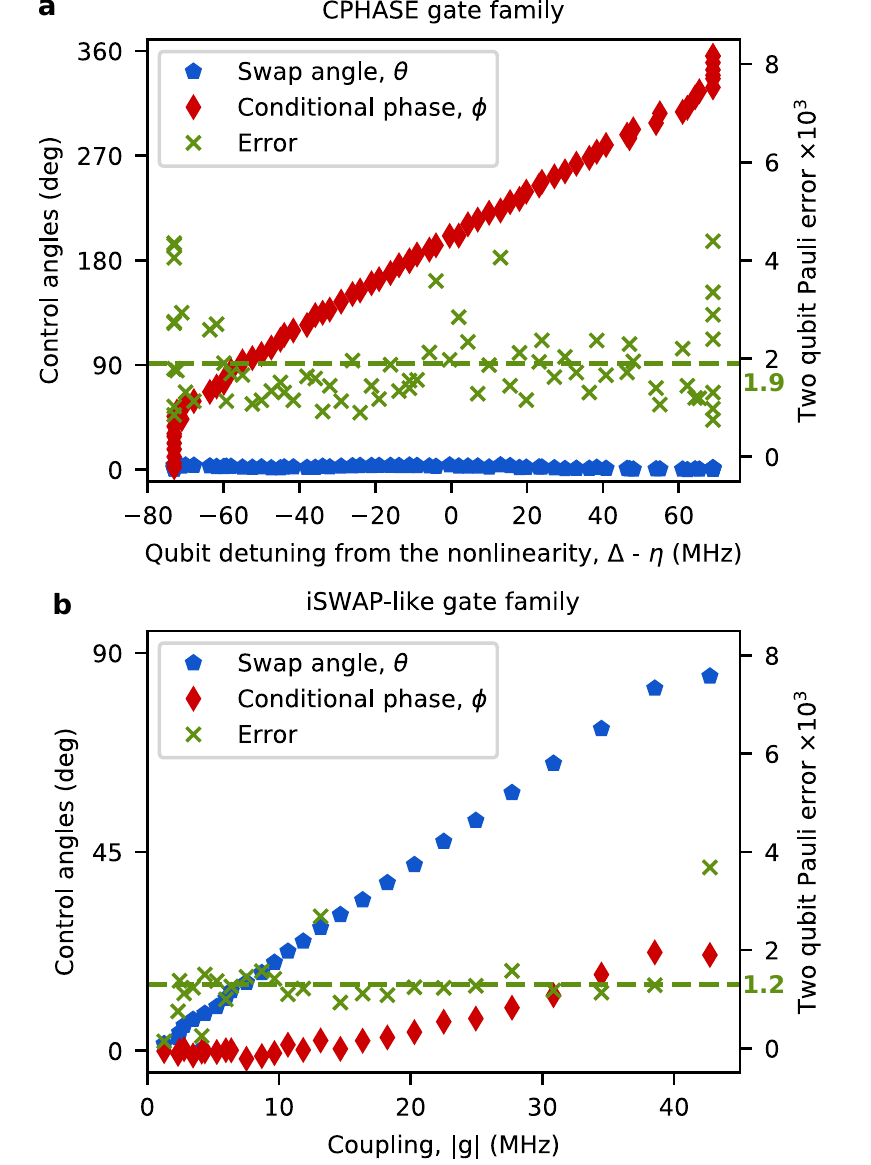}
\caption{\label{fig:line_cut_fidelity}Characterizing the iSWAP-like and CPHASE gate families with cross-entropy benchmarking. We plot the optimized fSim control angles, $\theta$ and $\phi$, on the left y-axes and the Pauli gate error per two-qubit gate on the right y-axes, conservatively assuming $\rm 7.5 \times 10^{-4}$ single-qubit Pauli gate errors. \textbf{a}, Characterization of the CPHASE gate family corresponding to $\rm fSim(\theta \approx 0^{\circ}, \phi)$. Each gate is 15\,ns long, consisting of control pulses that vary the qubit detuning, $\Delta$, around the qubit nonlinearity, $\eta$, with a coupler bias amplitude chosen to complete one full swap: $\ket{11} \rightarrow \ket{02} \rightarrow \ket{11}$. We measure an average two-qubit Pauli error of $\rm 1.9 \times 10^{-3}$ for the CPHASE family. \textbf{b}, Characterization of the iSWAP-like gate family corresponding to $\rm fSim(\theta, \phi \propto \theta^{2})$.  Each gate is 13\,ns long, consisting of control pulses that place the qubits on resonance and vary the coupling strength, $|g|$, to achieve an arbitrary swap angle $\theta$ between the $\ket{01}$ and $\ket{10}$ states. We measure an average two-qubit Pauli error of $\rm 1.2 \times 10^{-3}$ for the iSWAP-like family.}
\end{figure}

The data presented in Figure \ref{fig:background_scans} provides a map for implementing an arbitrary fSim\textemdash each pixel defines a set of three control amplitudes, and any control amplitudes yielding low-leakage should result in a high-fidelity gate described by the fSim control model (Eq. \ref{eq:fsim_main}). While it may be possible to perform an arbitrary fSim gate with a single set of flux pulses using either very strong coupling or more complex control waveforms, we have chosen to implement an arbitrary fSim gate as a composition of two continuous gate families using simple rectangular control pulses to minimize the gate length. The first gate family completes a diabatic $\ket{11} \rightleftarrows \ket{02}$ swap to perform a gate with an arbitrary conditional phase, $\phi$, using control amplitudes denoted by the dot-dashed line labeled 'CPHASE' in Figures \ref{fig:background_scans}a and \ref{fig:background_scans}b. The dominant control angle in the CPHASE gate family is the conditional phase, but, we do accumulate a small swap angle $\theta$ due to the strong coupling necessary to perform a fast CPHASE gate ($\theta \leq 5^{\circ}$ for a 13\,ns CPHASE gate\textemdash this may be reduced by increasing the gate duration). The second gate family places the qubits on resonance ($\Delta = 0\,$MHz) and varies $g$ to reach the desired swap angle, $\theta$, using control amplitudes along the dashed line labeled ``iSWAP-like'' in Figure \ref{fig:background_scans}c. We have deemed this gate family ``iSWAP-like'' since the swap angle varies from $\theta:[0^{\circ}, 90^{\circ}]$ and because this gate accumulates a conditional phase $\phi \propto \theta^2$ due to the dispersive interaction with the $\ket{02}$ and $\ket{20}$ states. Both of these gates are a subset of the fSim group individually, and, compiled together, they can reach the full fSim parameter space.

In Figure \ref{fig:line_cut_fidelity} we characterize both the iSWAP-like and CPHASE gate families using cross-entropy benchmarking (XEB) \cite{Boixo2018}. On the left axes we plot the optimized values of $\theta$ and $\phi$ for a range of CPHASE and iSWAP-like gates, and on the right y-axes we plot the Pauli error per two-qubit gate (see supplemental \ref{app:T1}), achieving average errors of $1.9  \times 10^{-3}$ and $1.2 \times 10^{-3}$ for each gate family respectively.

\section{\label{sec:fSim_fidelity} Benchmarking \lowercase{f}S\lowercase{sim} gates}

\begin{figure}
\includegraphics[width=89mm]{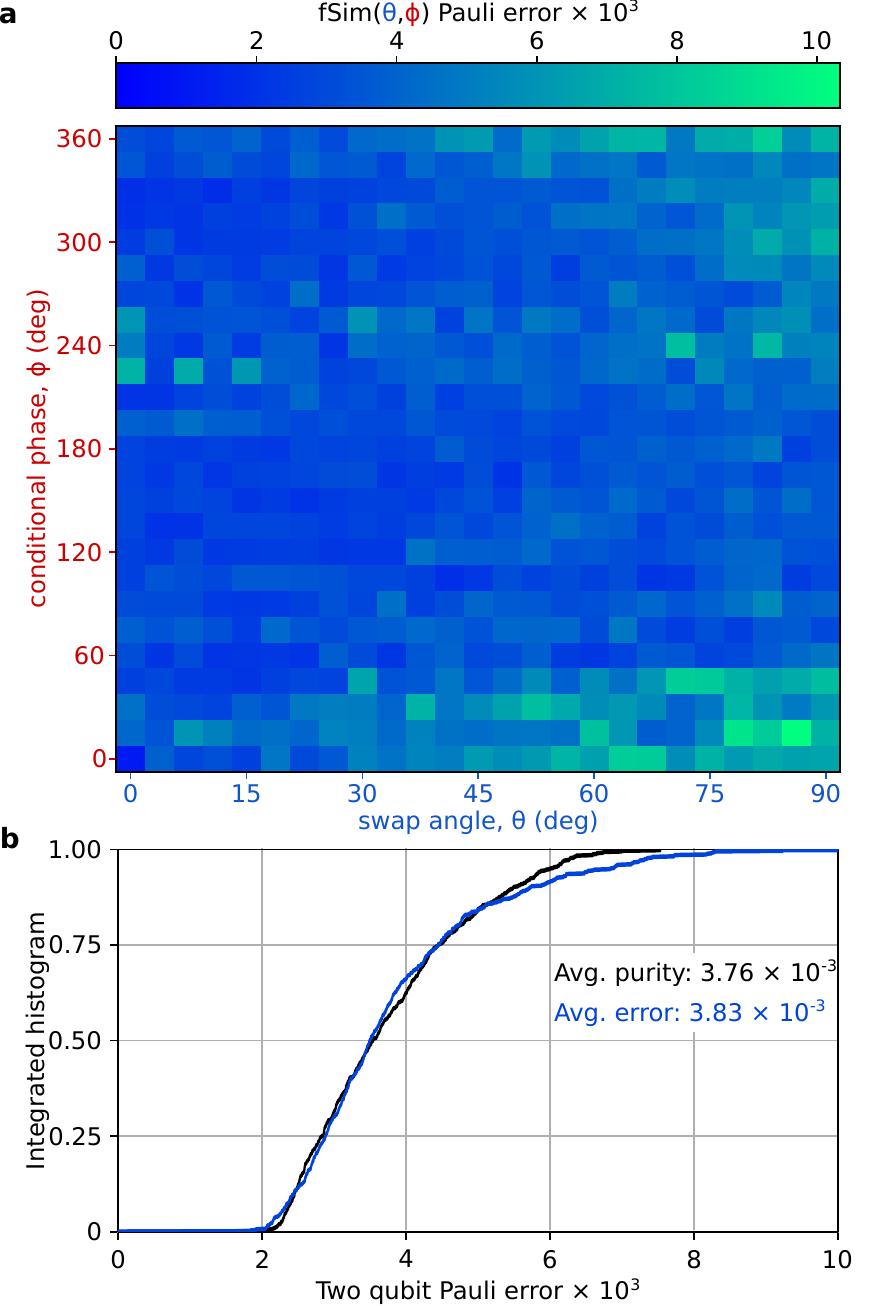}
\caption{\label{fig:fidelity} Benchmarking the fSim gate set. Using XEB, we measure the Pauli error per cycle and subtract off a conservative estimate for the single-qubit Pauli gate errors of $\rm 7.5 \times 10^{-4}$. \textbf{a}, We plot the two-qubit Pauli error for 525 fSim gates where $\rm \theta$ and $\rm \phi$ have been constrained to a grid. \textbf{b}, Histogram of both the error and purity for the gates presented in \textbf{a}. Here we confirm a purity (coherence) limited average error for our fSim gates of $\rm 3.83 \times 10^{-3}$.}
\end{figure}

In Figure \ref{fig:fidelity}a we present the Pauli error of 525 distinct fSim($\theta, \phi$) gates where the values of $\theta$ and $\phi$ have been constrained to be exactly the values indicated by the xy-coordinates at the center of each pixel (where \textit{ex situ} optimization has been used only to optimize the single-qubit phases). Each 28\,ns long fSim gate in Figure \ref{fig:fidelity} is a composition of a 15\,ns CPHASE gate followed by a 13\,ns iSWAP-like gate. While the fSim fidelity is largely independent of the values of $\theta$ and $\phi$ there are a few features of note. As discussed in supplement \ref{app:fsim_calibration}, we most-directly calibrated line cuts at $\theta = 0^{\circ}, 90^{\circ}$ and $\phi = 180^{\circ}$.  The regions of higher error where $\phi$ is near $0^{\circ}$ ($360^{\circ}$) involve the most extrapolation from the directly calibrated control amplitudes. Secondly, there is a faintly-visible indication of a band of higher error near $\phi \approx 240^{\circ}$ which we believe is due to a weakly interacting TLS defect in the spectrum of one of the qubits\textemdash in the future we hope to avoid such defects by shifting the frequencies of both qubits while maintaining their relative detuning. In Figure \ref{fig:fidelity}b we histogram these results in addition to the purity \cite{Wallman2015} per fSim and confirm a purity-limited average Pauli error of $\rm 3.83 \times 10^{-3}$ per fSim gate.

\section{\label{sec:conclusions}Conclusions}
We have implemented continuous iSWAP-like and CPHASE gate families with average Pauli error rates of $1.2 \times 10^{-3}$ and $1.9 \times 10^{-3}$ respectively. These fast (13-15\,ns) gates take advantage of the strong, tunable, qubit-qubit coupling offered by our gmon transmon qubit architecture achieving error rates more than a factor of two lower than the best previously reported two-qubit gates for superconducting qubits \cite{Barends2019}. Additionally, we have combined these two gate sets to demonstrate a complete implementation of the two-qubit fSim gate set with an average Pauli error of $3.83 \times 10^{-3}$ per gate. This direct implementation of the fSim gate offers roughly an additional factor of three in compilation efficiency for NISQ algorithms over a minimally-universal gate set.

\begin{acknowledgments}
This work was supported by Google LLC.  The UC Santa Barbara Nanofabrication Facility, part of the National Nanotechnology Infrastructure Network funded by NSF, fabricated the gmon qubits.
\end{acknowledgments}

$^*$ These authors contributed equally to this work

\bibliographystyle{unsrt}
\bibliography{apssamp}

\section{\label{app:fSim_parameterization}fSim control model}

\def\Id{{\rm Id}}
\def\CZ{{\rm CZ}}
\def\CR{{\rm CR}}
\def\iSWAP{{\rm iSWAP}}
\def\fSim{{\rm fSim}}
\def\T{{\rm T}}
\def\I{{\rm I}}
\def\X{{\rm X}}
\def\Y{{\rm Y}}
\def\Z{{\rm Z}}
\def\H{{\rm H}}
\def\rI{{\rm I}}
\def\rP{{\rm P}}
\def\A{{\cal A}}
\def\IQC{{\rm IQC}}
\def\bP{{\mathbb P}}
\def\cL{{\mathcal L}}

A generic representation of a Fermionic Simulation (fSim) gate corresponding to a two-qubit photon conserving unitary requires five parameters. We may separate out the single and two-qubit parameters as follows: a $\ket{01} \leftrightarrow \ket{10}$ swap angle, $\theta$, a $\ket{11}$ state conditional phase, $\phi$, and three single qubit phases, $\Delta_+, \Delta_-,$ and $\Delta_{-,off}$ yielding a generic fSim parameterization, 

\footnotesize
\begin{align}
\label{eq:fsim_full}
{\rm fSim}&(\theta,\phi, \Delta_+, \Delta_-, \Delta_{-,off} ) = \nonumber \\   
 &\begin{pmatrix}1 & 0 & 0 & 0 \\
 0 & e^{i (\Delta_+ + \Delta_-)} \cos \theta & -i e^{i (\Delta_+ - \Delta_{-,off})} \sin \theta & 0 \\
 0 & -i e^{i (\Delta_+ + \Delta_{-,off})} \sin \theta & e^{i (\Delta_+ - \Delta_-)} \cos \theta & 0 \\
 0 & 0 & 0 & e^{i (2\Delta_+ + \phi)} \\
\end{pmatrix}
\end{align}
\normalsize

We are interested in performing a two-qubit gate,
which is independent of the single-qubit rotations. Therefore, we
can focus on the matrix where $\Delta_+, \Delta_-,$ and $\Delta_{-,off}$are all zero, leading to the notation, 
\begin{align}
  \label{eq:19}
  {\rm fSim}(\theta,\phi) =  \begin{pmatrix}
    1 & 0 & 0 & 0 \\
    0 & \cos \theta &  -i \sin \theta & 0 \\
    0 & -i \sin \theta & \cos \theta & 0 \\
    0 & 0 & 0 & e^{-i \phi} \\
  \end{pmatrix}
\end{align}
used to designate an arbitrary gate within the excitation preserving subspace. 

\section{\label{app:numerics}fSim gate numerics}

\begin{figure*}
\includegraphics[width=163mm]{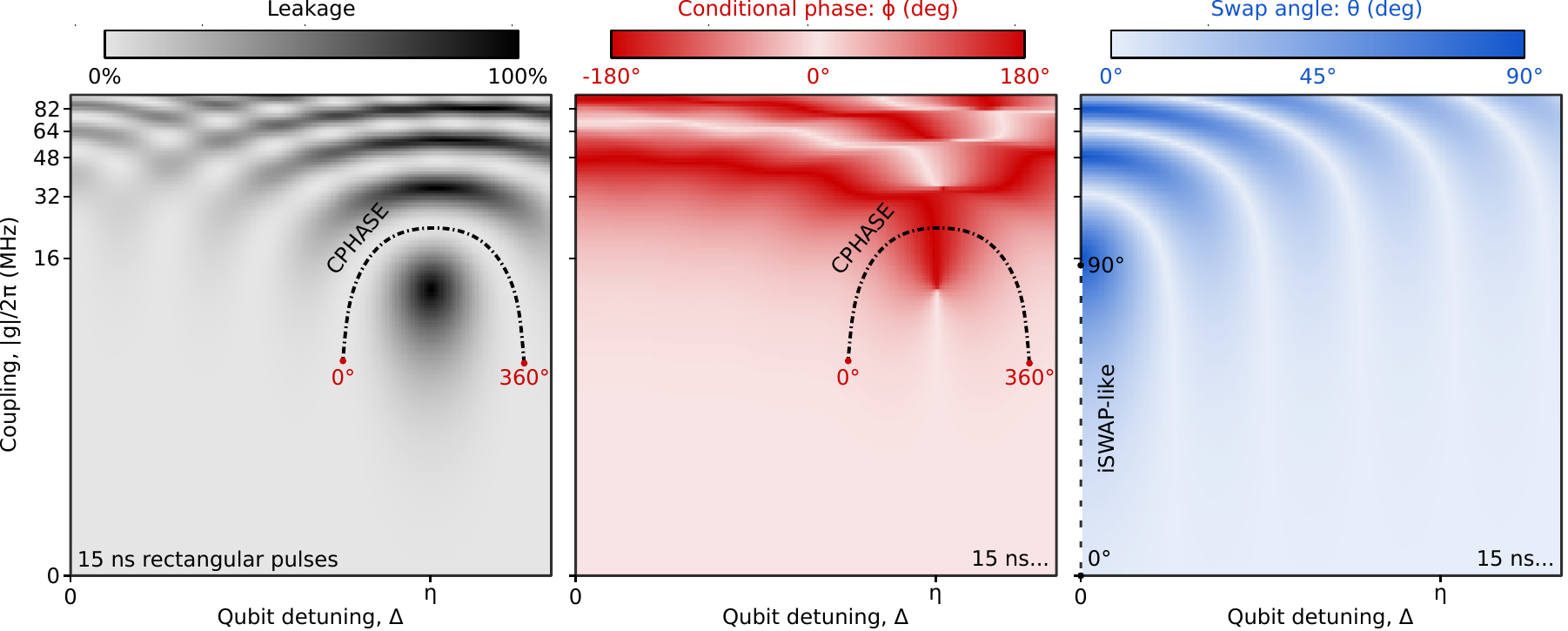} \caption{
Numeric simulation of two interacting qutrits reproducing the data from our experiments in Figure \ref{fig:background_scans} of the main text. We simulate qubits with a fixed nonlinearity (240\,MHz) with 15\,ns long rectangular control pulses defining the qubit detuning, $\Delta$, and their coupling, $g$.
\label{fig:sim_background_scans}}
\end{figure*}

The qubit dynamics presented in the main paper (Figure \ref{fig:background_scans}) are well described by numerics simulating two interacting qutrits (e.g. a pair of coupled three-level anharmonic oscillators) evolving with a time dependent detuning, $\Delta(t)$, and coupling, $g(t)$. We truncate the full two-qutrit Hamiltonian limiting our simulation to states with 1 or 2 excitations. Operating with the basis $\ket{01}, \ket{10}, \ket{11}, \ket{20}, \ket{02}$, the Hamiltonian describing the system is given by:

\begin{eqnarray}
\rm H(g, \Delta, \eta) = \left(\begin{array}{ccccc}
     0 & g & 0 & 0 & 0 \\
     g & \Delta & 0 & 0 & 0 \\
     0 & 0 & \Delta & \sqrt2 g & \sqrt2 g \\
     0 & 0 & \sqrt2 g & 2\Delta + \eta & 0\\
     0 & 0 & \sqrt2 g & 0 & \eta \\
\end{array}\right)
\end{eqnarray}

\noindent where $\eta$ is the nonlinearity of each qubit, which we assume is the same for both qubits (240\,MHz). Using this model, we may estimate the unitary operation enacted by arbitrary time-domain control of the coupling strength and the qubit detuning by discretizing these time domain control waveforms and performing a time ordered integral of $\rm H(t)$.

\begin{figure*}
\includegraphics[width=163mm]{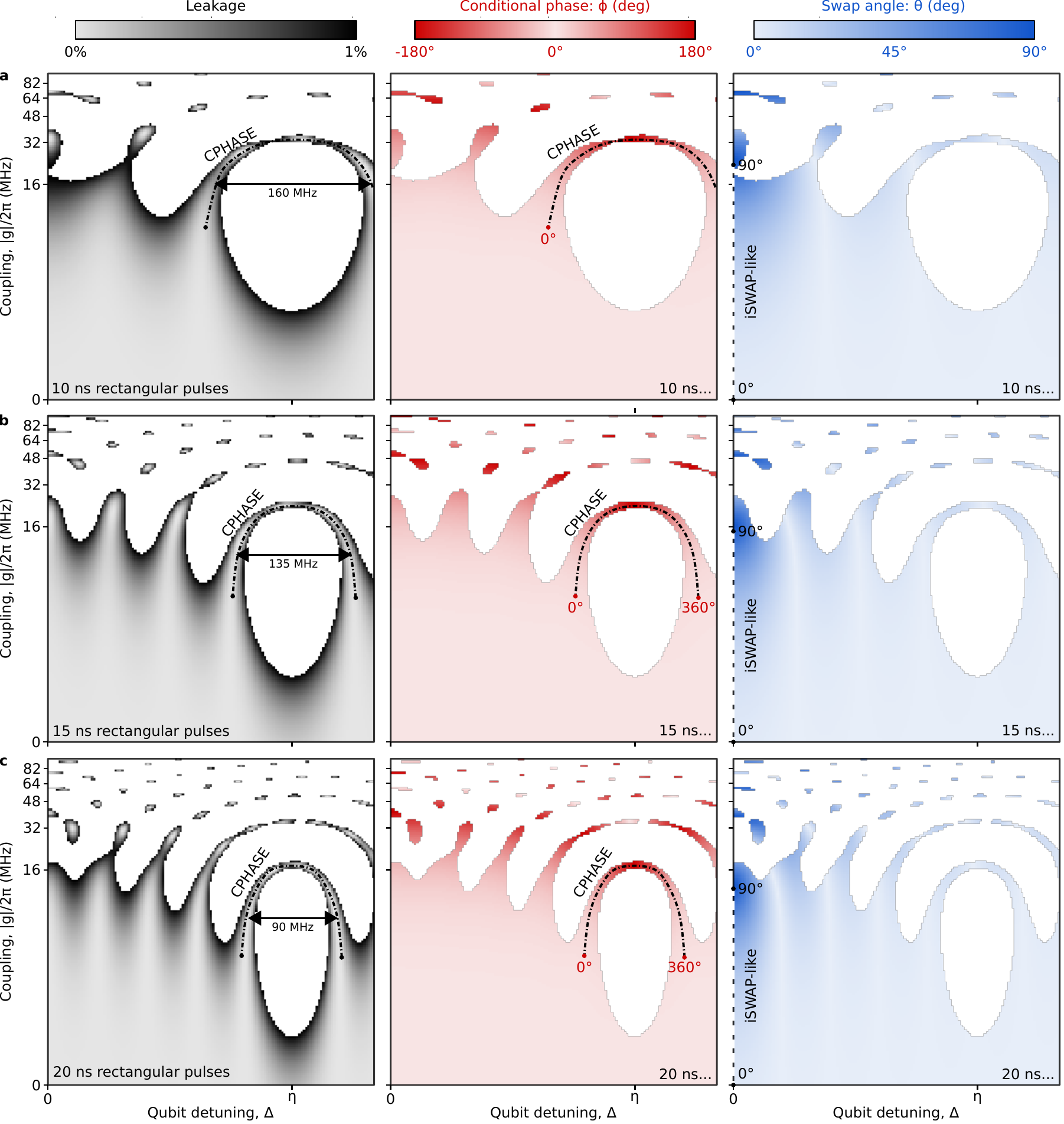} \caption{
Numeric simulation of \textbf{a}, 10\,ns, \textbf{b}, 15\,ns, and \textbf{c}, 20\,ns rectangular control pulses showing the fSim parameter space where leakage is less than 1\% (white regions are where leakage exceeded this threshold). Experimentally we chose to perform our CPHASE gate with 13\,ns long pulses and the iSWAP-like gate with 11\,ns control pulses (both of which had 1\,ns pads on either side)\textemdash as we found that shorter implementations of either gate increased leakage and the overall gate error. Here, these numerics demonstrate that for 10\,ns long gates, the low-leakage lobe where we perform the CPHASE gate narrows considerably and the $\ket{2}$ state Rabi interaction reaches the on-resonance iSWAP-like line cut near $\rm \theta = 90^{\circ}$, both of which agree with our experimental results.
\label{fig:sim_length}}
\end{figure*}

\begin{figure*}
\includegraphics[width=163mm]{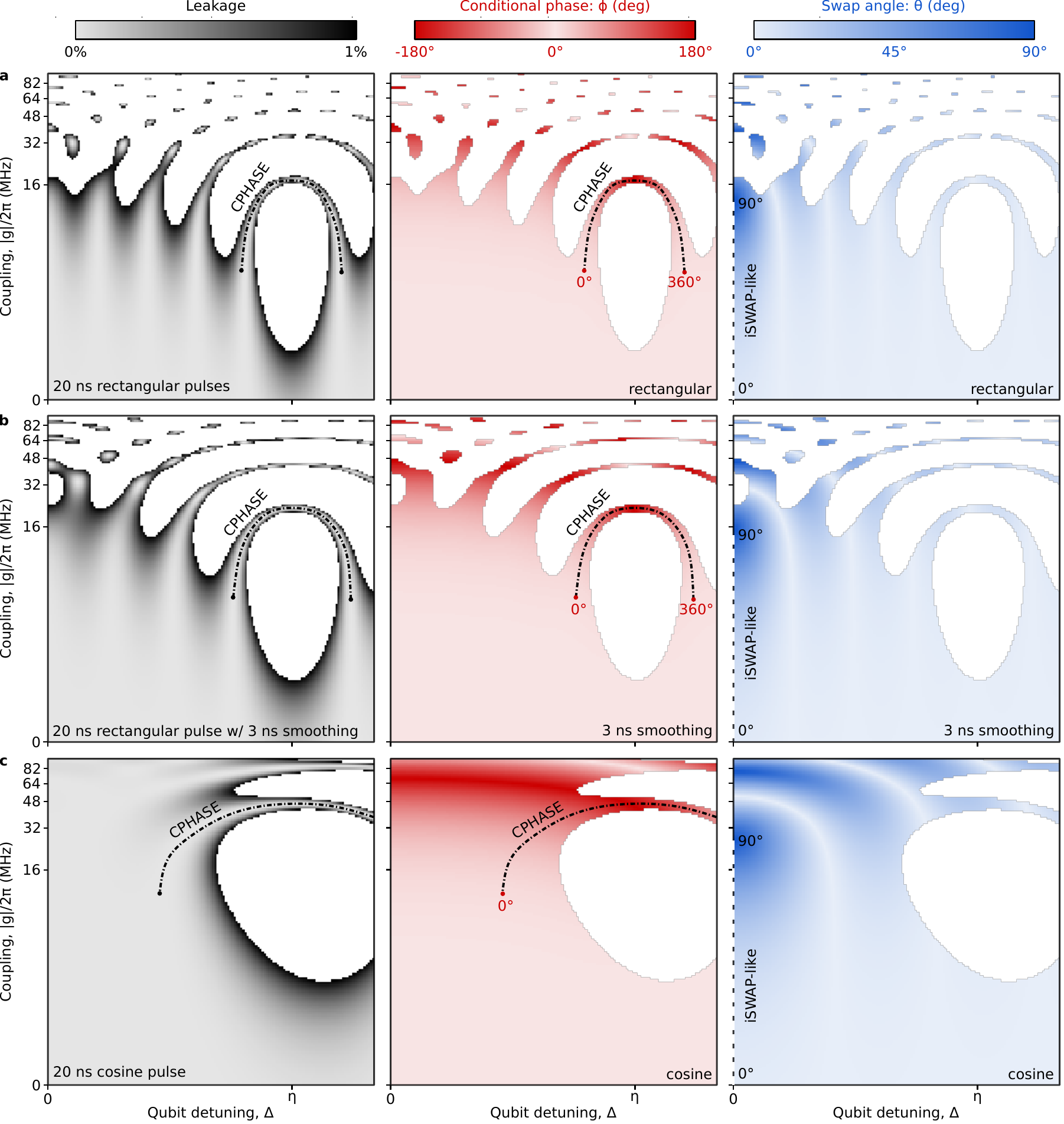} \caption{
Numeric simulation of \textbf{a}, a 20\,ns rectangular coupler pulse, \textbf{b}, a 3\,ns rise time rectangular pulse, and \textbf{c}, cosine coupler pulse showing the fSim parameter space where leakage is less than 1\%. We observe that as the coupler pulses become more smooth, the fSim parameters space where leakage is less than 1\% expands considerably.  This indicated that pulse shaping and or smoothing may play an important role in any future implementation of the fSim gate set that aims to implement the gate set with a single pulse.
\label{fig:sim_smooth}}
\end{figure*}

In Figure \ref{fig:sim_background_scans} we qualitatively reproduce the experimental results in Figure \ref{fig:background_scans} by simulating 15\,ns rectangular control pulses defining both $g$ and $\rm \Delta$. In Figure \ref{fig:sim_length} we illustrate the broadening effect that using shorter pulse lengths has on the Rabi interactions of both the $\ket{01} \leftrightarrow \ket{10}$ and $\ket{11} \leftrightarrow \ket{02}$ interactions by simulating rectangular pulses that are 10\,ns, 15\,ns, and 20\,ns long. In Figures \ref{fig:sim_length} and \ref{fig:sim_smooth}, we have omitted points where the leakage exceeds a 1\% threshold which identifies the parameter space where we can perform fSim gates with low error. Experimentally we have chosen to implement our CPHASE gates with 13\,ns long rectangular pulses with a 1\,ns pad on either side\textemdash when we made the gate length shorter, leakage increased (data not shown). Here, in Figure \ref{fig:sim_length}a, we qualitatively see that the width of the 1\% leakage band where we perform the CPHASE gate begins to pinch off and the $\ket{2}$ state Rabi interaction reaches all the way to the on-resonance iSWAP-like parameter space (dotted white line) when the gate length is 10\,ns. Both these results qualitatively reproduce what we observed experimentally when attempting iSWAP-like gates shorter than 11\,ns or the CPHASE gate shorter than 13\,ns. Finally, in Figure \ref{fig:sim_smooth} we simulate the effect of smoothing the control pulses by simulating 20\,ns long coupler pulses that are rectangular, rectangular with 3\,ns Gaussian smoothing, and cosine shaped (all detuning pulses are rectangular and have the same length). Here we see that smoothing reduces the extent of leakage from the second and third $\ket{11} \leftrightarrow \ket{02}$ swap lobes expanding the available low-error fSim control space. This indicates that pulse smoothing may be an important consideration of any future fSim implementation that aims to perform an arbitrary fSim using a single coupler pulse instead of the two discrete rectangular pulses we have used in this work.

\newpage
\section{\label{app:gate_characterization}Gate characterization}
We use a variety of techniques to characterize the performance of our single and two-qubit gates which we detail in this section. In lieu of full process tomography, we use depth one population based measurements to perform unitary tomography to quickly assess the unitary operation performed by a given set of control pulses. We then turn to benchmarking techniques that amplify gate errors and allow for the characterization of small error rates. We use Clifford based benchmarking to characterize our single-qubit microwave gates and cross-entropy benchmarking (XEB) to characterize our two-qubit entangling gates.

\subsection{\label{app:Pauli_errors}Computing and reporting Pauli error rates}
Before jumping in to gate characterization, a quick aside on Pauli error rates. We report Pauli error rates which are independent of the Hilbert space dimension and thus add linearly as the circuit's Hilbert space grows. In the past, many have reported average single and two-qubit error, $e_r$, as exponential decay constants of a sequence fidelity, $F = A e^{me_r} + B$ where $A$ and $B$ are fit parameters to compensate for state preparation and measurement (SPAM) errors, $m$ is the number of gate repetitions in the sequence, and $e_r$ is the error per cycle. The Pauli error, $e_p$, is related to $e_r$ by the dimension of the Hilbert space:

\begin{equation}
    e_p = e_r \times \Big( 1 + \frac{1}{D} \Big)
\end{equation}

\noindent where $\rm D = 2^n$ is the dimension of the Hilbert space for an n-qubit gate. We note that this results in an increase in the reported error by a factor of 1.5 for single-qubit gates ($n=1$) and a by a factor of 1.25 for two-qubit gate errors ($n=2$).

When performing two-qubit XEB, we measure the exponential decay constant per cycle, $e_{r,\rm{cycle}}$ where each cycle consists of the application of one single-qubit gate per qubit and one fSim entangling gate involving both qubits. In order to extract the error per fSim gate, we can convert this to a Pauli error per cycle, $e_{p,\rm{cycle}}$, and subtract off the two single-qubit Pauli gate errors, $e_{p,\rm{q_1}}$ and $e_{p,\rm{q_2}}$, which we estimate using single-qubit Clifford based randomized benchmarking.

\begin{equation}
    e_{p,\rm{2q}} = e_{p,\rm{cycle}} - (e_{p,\rm{q_1}} + e_{p,\rm{q_2}})
\end{equation}

\noindent For simplicity, all two-qubit Pauli errors have been computed assuming single-qubit Pauli errors of $7.5 \times 10^{-4}$ per gate per qubit consistent with our typical single-qubit error rates immediately following a successful run of our standard single-qubit gate calibration procedure (see supplement \ref{app:T1}).

\begin{equation}
    e_{p,\rm{two\_qubit}} = e_{p,\rm{cycle}} - (2 \times 7.5 \times 10^{-4}) = e_{p,\rm{cycle}} - 1.5\times10^{-3}
\end{equation}

\newpage 
\subsection{\label{app:T1}Single-qubit coherence and gates}
\begin{figure}
\includegraphics[width=89mm]{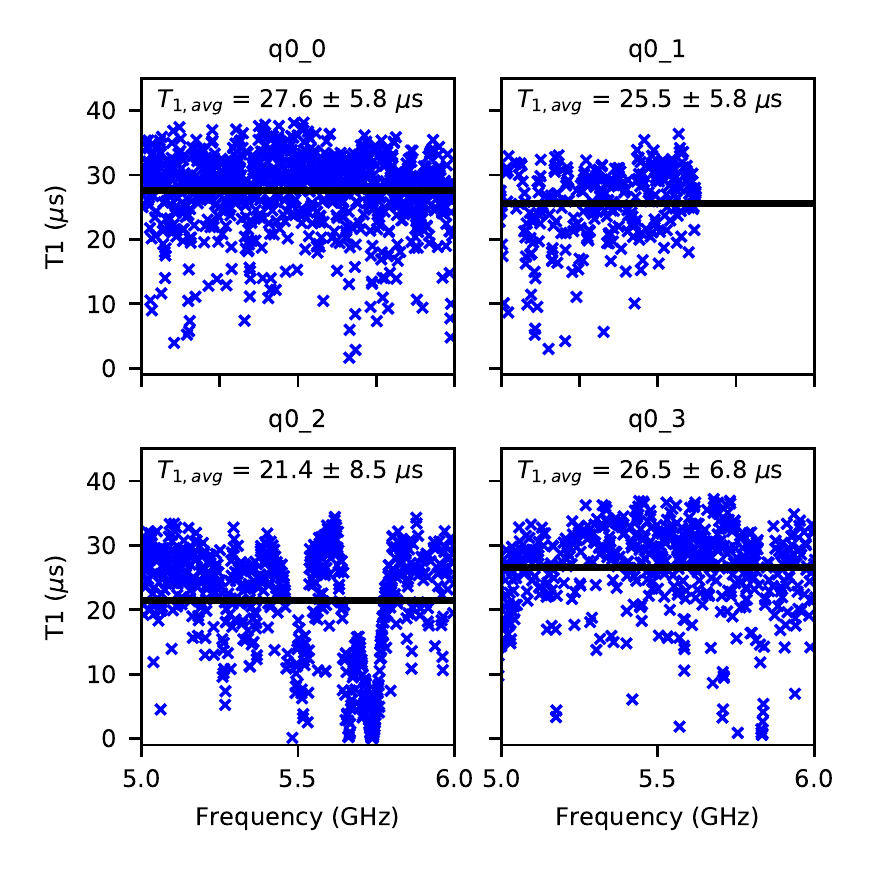}
\caption{\label{fig:T1}Swap spectroscopy for four qubits from 5 to 6\,GHz characterizing the qubit $T_1$ as a function of qubit frequency. For all four qubits on this chip over the available frequencies in the range of 5-6\,GHz we find an average $T_1 = 25.3 \pm 7.3 \mu$s.} 
\end{figure}

Qubit coherence, in conjunction with gate duration, places a lower bound on both our single and two-qubit gate error rates.  In Figure \ref{fig:T1} we characterize $T_1$ for four qubits over a frequency range of 5 to 6\,GHz. To perform this measurement we calibrate single-qubit gates, readout, and flux bias frequency control for a given qubit idle frequency. We then excite the qubit to the $\ket{1}$ state and detune the qubit to another frequency for a variable amount of time before detuning back to the idle frequency for readout. For each detuned frequency, $T_1$ is extracted as an exponential decay of the population over time, $P\ket{1} \propto A e^{-t/T_1} + B$, where $A$ and $B$ are fit parameters to compensate for state preparation and measurement errors. We find $T_1 = 25.3 \pm 7.3\,\mu$s averaging data from all four qubits over a frequency range of $\rm 5-6\,GHz$. Since $f_{\rm{max}}$ for the second qubit was anomalously low, we averaged data for this qubit from $\rm 5-5.61\,GHz$.  

\begin{figure}
\includegraphics[width=89mm]{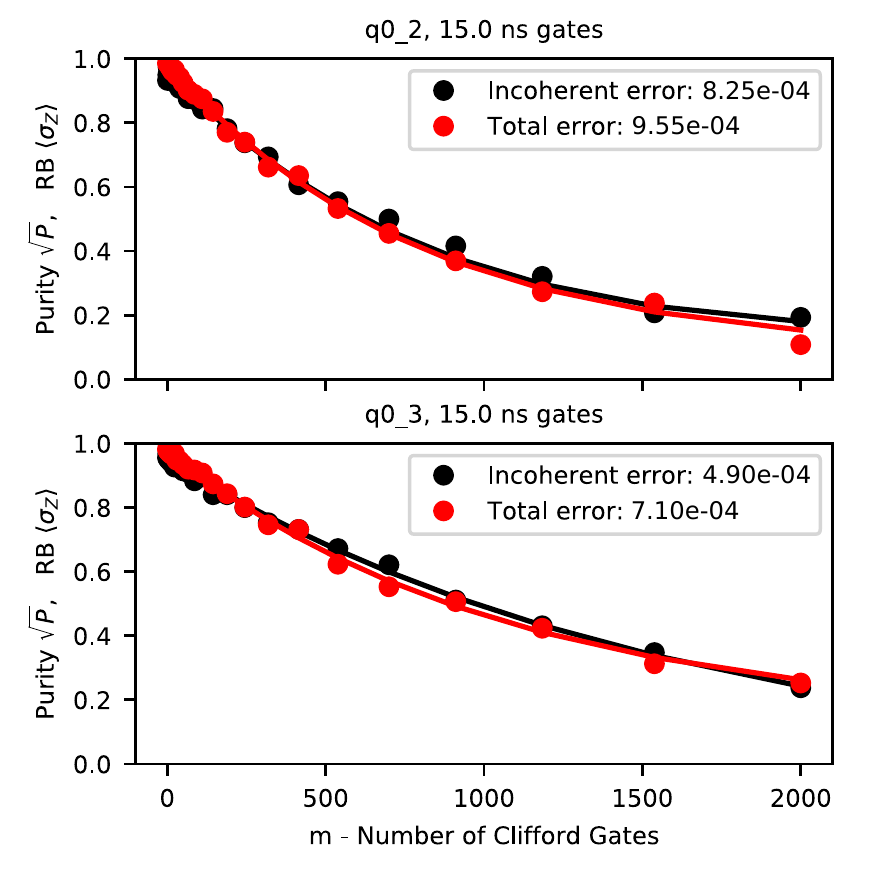}
\caption{\label{fig:sq_rb} Representative single-qubit Clifford-based randomized benchmarking results used to characterize the average error of our single-qubit gates. With a typical calibration, the single-qubit Pauli errors for both qubits are usually in the range of $5 - 10 \times 10^{-4}$. When computing the two-qubit gate error from the XEB per cycle error throughout this paper, we assume a moderately conservative error of $7.5 \times 10^{-4}$ per single-qubit gate. } 
\end{figure}

We use single-qubit purity \cite{Wallman2015} and Clifford-based randomized benchmarking \cite{Magesan2011, Corcoles2013} to characterize the average error of our single-qubit gates. In Figure \ref{fig:sq_rb} we present representative results for a pair of qubits demonstrating purity-limited (incoherent error-limited) performance. These gate errors drift over time, but immediately following a successful run of our standard calibration procedure we typically observe single-qubit error rates at or slightly higher than the $7.5 \times 10^{-4}$ level \cite{ZChen2016}.  As such, we use this estimate in computing two-qubit error rates throughout this paper. These error rates are consistent with the coherence limit, for $\rm T_{\rm{gate}} = 15$\,ns and $\rm T_1 = 30$\,$\rm \mu$s, giving $e_{p,\rm{inc}} \approx 1.5 \times \rm{T_{gate}}/3 \rm{T_1}) = 2.5 \times 10^{-4}$, with the remainder of the error coming from leakage and $\rm T_{2}$ \cite{O'Malley2015}.

\subsection{\label{app:unitary_tomography}Unitary tomography}
Section II of the main text describes shallow circuits used to characterize leakage and the two-qubit control parameters, $\theta$ and $\phi$.  Here, we detail the procedure used to directly measure all the non-zero matrix elements composing an arbitrary photon conserving unitary operation and the algebra used to convert these matrix elements into the five fSim control parameters (in Eq. \ref{eq:fsim_full}). We use the resulting fSim model to compute the XEB sequence fidelity which we may then use as a cost function to optimize some, or all, of the fSim model parameters. 

\begin{table}
\caption{\label{table:unitary_tomography}Summary of the two-qubit unitary tomography measurement sequences. Here, \{$u_{11}, u_{12}, u_{22}, u_{21}$\} are the complex matrix elements of the two-qubit unitary in [$\ket{01}, \ket{10}$] subspace. The two additional measurements ($u_{12,\rm{excited}}$ and $u_{22,\rm{excited}}$) are repeated measurements of $u_{12}$ and $u_{22}$ but with the other qubit placed into the excited state. This additional information is used to construct the conditional phase, $\phi$.}
\footnotesize

\begin{tabular}{|c|c|c|}
\hline
Matrix element & Initial state &  Measure qubit \\ \hline
$u_{11}$ & (x, 0) & 0 \\ \hline
$u_{12}$ & (0, x) & 0 \\ \hline
$u_{22}$ & (0, x) & 1 \\ \hline
$u_{21}$ & (x, 0) & 1 \\ \hline
$u_{21,\rm{excited}}$ & (1, x) & 0 \\ \hline
$u_{22,\rm{excited}}$ & (1, x) & 1 \\ \hline

\end{tabular}\\
\end{table}
\normalsize

In order to efficiently characterize the unitary operation performed by a given set of control pulses, we initialize and measure a set of circuits as summarized in Table \ref{table:unitary_tomography}. If we consider a general photon conserving unitary the non-zero matrix elements will take the form:

\begin{eqnarray}
\rm U = {\begin{array}{cccc}
     \ket{00} & \ket{01} & \ket{10} & \ket{11}
     \end{array} \atop
     \left(\begin{array}{cccc}
      1 & 0 & 0 & 0  \\
     0 & u_{11} & u_{12} & 0 \\
     0 & u_{21}  & u_{22} & 0 \\
     0 & 0 & 0 & u_{33} 
     
\end{array}\right)}
{\begin{array}{c}
     \\
     \end{array} \atop
     \begin{array}{c}
      \ket{00} \\
      \ket{01} \\ 
      \ket{10} \\
      \ket{11}
      \end{array}}
\end{eqnarray}

Where $u_{nm}$ denotes a non-zero element. We measured $u_{nm}$ by initializing excited qubit in the basis ket of column $m$ with an $X/2$ gate, and measuring the expectation value of $\sigma_x + i \sigma_y$ of the excited qubit in the basis ket denoted by row $n$. e.g. for $u_{21}$ we initialize the left qubit, apply the fSim gate, and then measure $\sigma_x + i \sigma_y$ of the right qubit\textemdash this is the complex value of $u_{21}$. This procedure works for the single excitation subspace (e.g. $n,m$ in $[1,2]$), but $u_{33}$ is computed from repeated measurements of $u_{12,\rm{excited}}$ and $u_{22,\rm{excited}}$ where the previously uninitialized qubit is instead placed into the $\ket{1}$ state as summarized in Table \ref{table:unitary_tomography}. This procedure is similar to process tomography, but requires considerably fewer measurements to characterize the fSim matrix. We note that an optimal measurement sequence would require only 2$n$-1 circuits (for a $n \times n$ matrix) \cite{Rahimi-Keshari2013}.  Even with several thousand repetitions of each circuit, characterizing the matrix with this method takes only a few seconds. Our series of six circuits is intentionally over-complete to avoid singular behavior when some matrix elements are small. In table \ref{table:fsim_from_unitary} we list the conversion matrix elements to the five parameters of our fSim control model. These are useful measurements for building an fSim model, but we cannot characterize small gate errors ($\approx 10^{-3}$) using this method due to the limitations of state preparation and measurement (SPAM) errors which are a few percent.

\begin{table}
\caption{\label{table:fsim_from_unitary}Computing fSim model parameters from the results of our unitary tomography protocol. The ``condition'' column is present because we compute $u_{33} = u_{22,\rm{excited}}/u_{11}^*$ or $u_{33} = u_{12,\rm{excited}}/u_{21}^*$ depending on if $u_{11}$ or $u_{21}$ is larger to ensure the result is non-singular. $\psi_{10}$ is  the phase difference accumulated between the two qubits over the gate duration.}
\scriptsize

\begin{tabular}{|c|c|c|}
\hline
fSim parameter & Value &  condition \\ \hline
$\theta$ & $\arctan(|u_{12}|/|u_{11}|)$ & none \\ \hline
$\phi$ & $\Delta_+ - \angle(u_{12,\rm{excited}} \times u_{21})$ & $|u_{21}|>|u_{11}|$ \\ \hline
$\phi$ & $\angle(u_{22}) - \angle(u_{22,\rm{excited}})$ & $|u_{21}|<|u_{11}|$ \\ \hline
$\Delta_+$ & $\angle(-u_{11}\times u_{21})$ & $|u_{21}|>|u_{11}|$ \\ \hline
$\Delta_+$ & $\angle(u_{11}\times u_{22})$ & $|u_{21}|<|u_{11}|$ \\ \hline
$\Delta_-$ & $2 \times \angle(u_{11})- \Delta_+$ & none \\ \hline
$\Delta_{-,\rm{off}}$ & $-2\,(\angle(-u_{12}/\dot{\imath}) + \psi_{10})  + \Delta_+$ & $\psi_{10} = (\omega_{q_1} - \omega_{q_0}) * t_{gate}$ \\ \hline
\end{tabular}\\

\end{table}
\normalsize

\subsection{\label{app:xeb}Cross-entropy error benchmarking}
Cross-entropy benchmarking (XEB) is a powerful technique for characterizing the error of an arbitrary gate \cite{Boixo2018}. It is particularly useful when implementing non-Clifford gates like the continuous fSim gate set we use here. XEB uses a repetitive gate sequence to amplify small errors where each cycle consists of a random single-qubit gate from the set \{X/2, Y/2, $\pm$X/2$\pm$Y/2\} applied to each qubit followed by the fSim gate we are benchmarking. We extract the error per cycle as an exponential decay in the XEB sequence fidelity, $\mathcal{F}_{\rm{XEB}}$.  The sequence fidelity is computed using the cross-entropy between two probability distributions $P$ and $Q$, $S(P, Q) = -\sum_i p_i ln(q_i)$, by comparing the expected, measured, and incoherent probability distributions for a given gate sequence,

\begin{equation}
   \mathcal{F}_{\rm{XEB}}= \frac{S(P_{\rm incoherent}, P_{\rm expected}) - S(P_{\rm measured}, P_{\rm expected})}{S(P_{\rm incoherent}, P_{\rm expected}) - S(P_{\rm expected})  }
\end{equation}

\noindent The numerator is the difference between the measured and expected cross-entropy and the denominator serves as a normalization so that $\mathcal{F}_{\rm{XEB}}$ takes a value from [0, 1]. We then use $1-\mathcal{F}_{\rm{XEB}}$ as a cost function to optimize the five parameters of our fSim control model. For a given random sequence, we compute the expected probability distribution using perfect single-qubit gate models and the fSim model obtained from our unitary tomography experiment (supplement \ref{app:unitary_tomography}). Since, the sequence fidelity is dependent on the single and two-qubit gate models used in the cross-entropy calculation, we can use $1- \mathcal{F}_{\rm{XEB}}$ as a cost function to optimize some or all of our fSim gate model parameters, a process termed \textit{ex\,situ} optimization.

\subsection{\label{app:rb_vs_xeb}RB vs XEB}

\begin{figure}
\includegraphics[width=89mm]{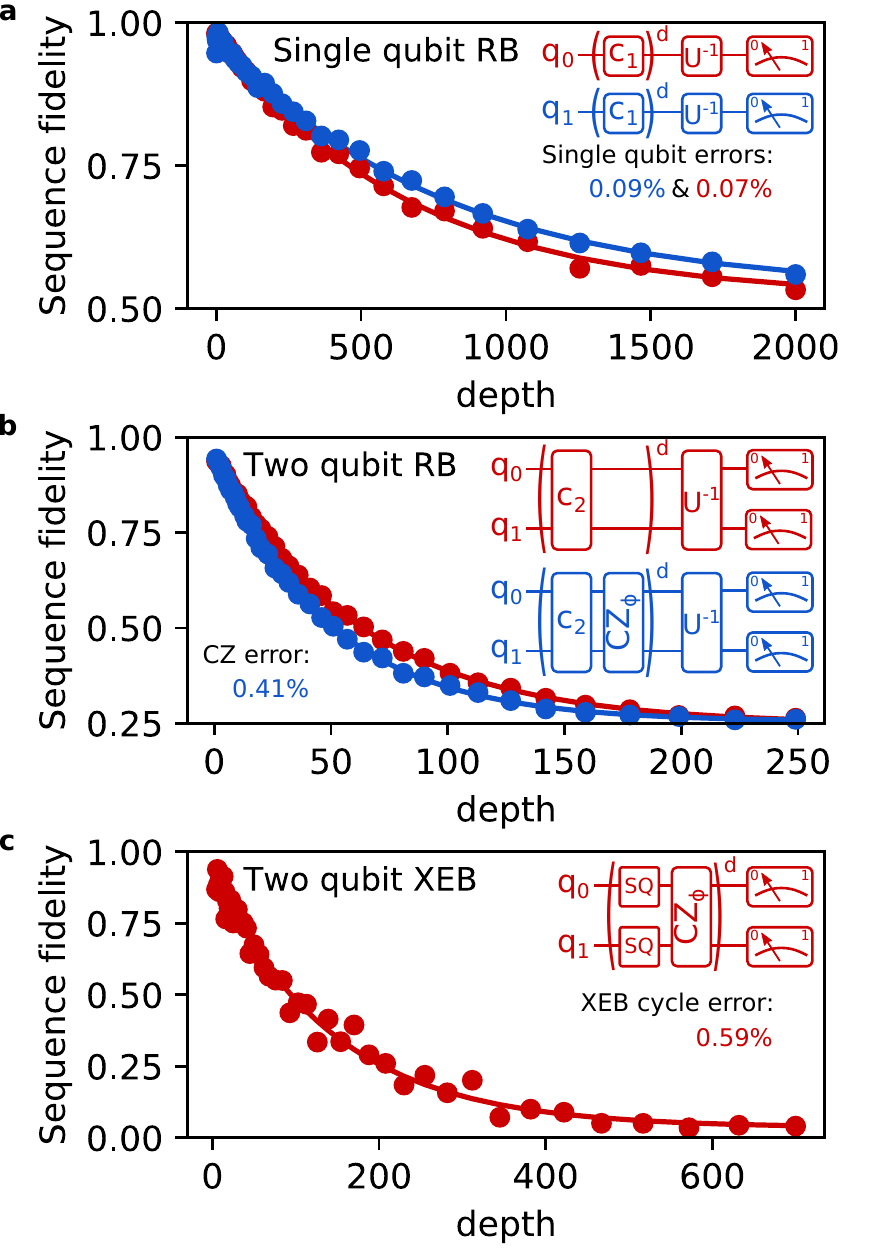}
\caption{\label{fig:rb_vs_XEB} Comparison of Clifford-based randomized benchmarking (RB) and cross-entropy benchmarking (XEB). \textbf{a}, Single-qubit Clifford based randomized benchmarking measuring average Pauli errors of 0.09\% and 0.07\% for each qubit. \textbf{b}, Two-qubit Clifford based randomized benchmarking with (blue) and without (red) an interleaved $\rm CZ_{\phi}$ gate, allowing us to extract the Pauli error per $\rm CZ+{\phi}$ of 0.41\%. \textbf{c}, Two-qubit cross-entropy benchmarking where each cycle includes two single-qubit gates and a $\rm CZ_{\phi}$ gate yielding a Pauli error per cycle of 0.59\%. Here we find that the sum of the single and two-qubit errors measured with Clifford based RB (0.09\% + 0.07\% + 0.41\% = 0.57\%) corresponds well to the XEB error per cycle (0.59\%).}
\end{figure}

As a sanity check, one may ask that we compare the result of Clifford based randomized benchmarking (RB) and cross-entropy benchmarking (XEB). Clifford based RB requires an inversion gate, inverting a random gate sequence to map the total ideal gate sequence starting in the $\ket{0}$ state back to $\ket{0}$. For most of the fSim gates, the inversion gate is non-trivial, but, for the special case of a $\rm CZ_{\phi} = fSim(0^{\circ}, 180^{\circ})$, which is part of the Clifford gate set, this comparison is possible.

In Figure \ref{fig:rb_vs_XEB}a we perform single-qubit Clifford based randomized benchmarking (gate sequence inset), extracting average single-qubit Pauli errors $e_{p,q1} = 0.7 \times 10^{-3}$ and $e_{p,q2} = 0.9 \times 10^{-3}$. In Figure \ref{fig:rb_vs_XEB}b we perform two-qubit Clifford based randomized benchmarking with and without an interleaved $\rm CZ_{\phi}$ gate (sequences inset), extracting a Pauli error per $\rm CZ_{\phi}$ of $4.1 \times 10^{-3}$. Then, in Figure \ref{fig:rb_vs_XEB}c we use XEB to measure the per cycle error of the $\rm CZ_{\phi}$ + two single-qubit gates obtaining $e_{p,\rm{cycle}} = 5.7 \times 10^{-3}$.  If we then sum the Clifford based errors for each SQ gate and the $\rm CZ_{\phi}$ $(0.7 + 0.9 + 4.1) \times 10^{-3} = 5.7 \times 10^{-3}$ we find good agreement with the XEB error per cycle $e_{p,\rm{cycle}} = 5.9 \times 10^{-3}$.

\subsection{\label{app:error_budget}Error budgeting}
In this section, we use various techniques to provide a more thorough budget of our XEB per cycle errors. As we have discussed, XEB measures the total error per cycle, $e_{p,\rm{cycle}}$. This includes coherent and incoherent errors for one single-qubit gate per qubit and one fSim gate. We use single-qubit Clifford-based randomized benchmarking to characterize the average total error for single-qubit gates, we use purity benchmarking to characterize incoherent error of both the single-qubit and fSim gates, and we use $\ket{2}$ state readout in conjunction with XEB to characterize per cycle leakage (which is included in the incoherent error). Here we focus on the two-qubit gate errors by assuming purity-limited single-qubit Pauli gate errors of $7.5 \times 10^{-4}$ as described in supplement \ref{app:T1}\textemdash this effectively means we subtract $1.5 \times 10^{-3}$ from $e_{p,\rm{cycle}}$ to obtain $e_{p,\rm{2q}}$ for both error and purity measurements..

\begin{figure*}
\includegraphics[width=183mm]{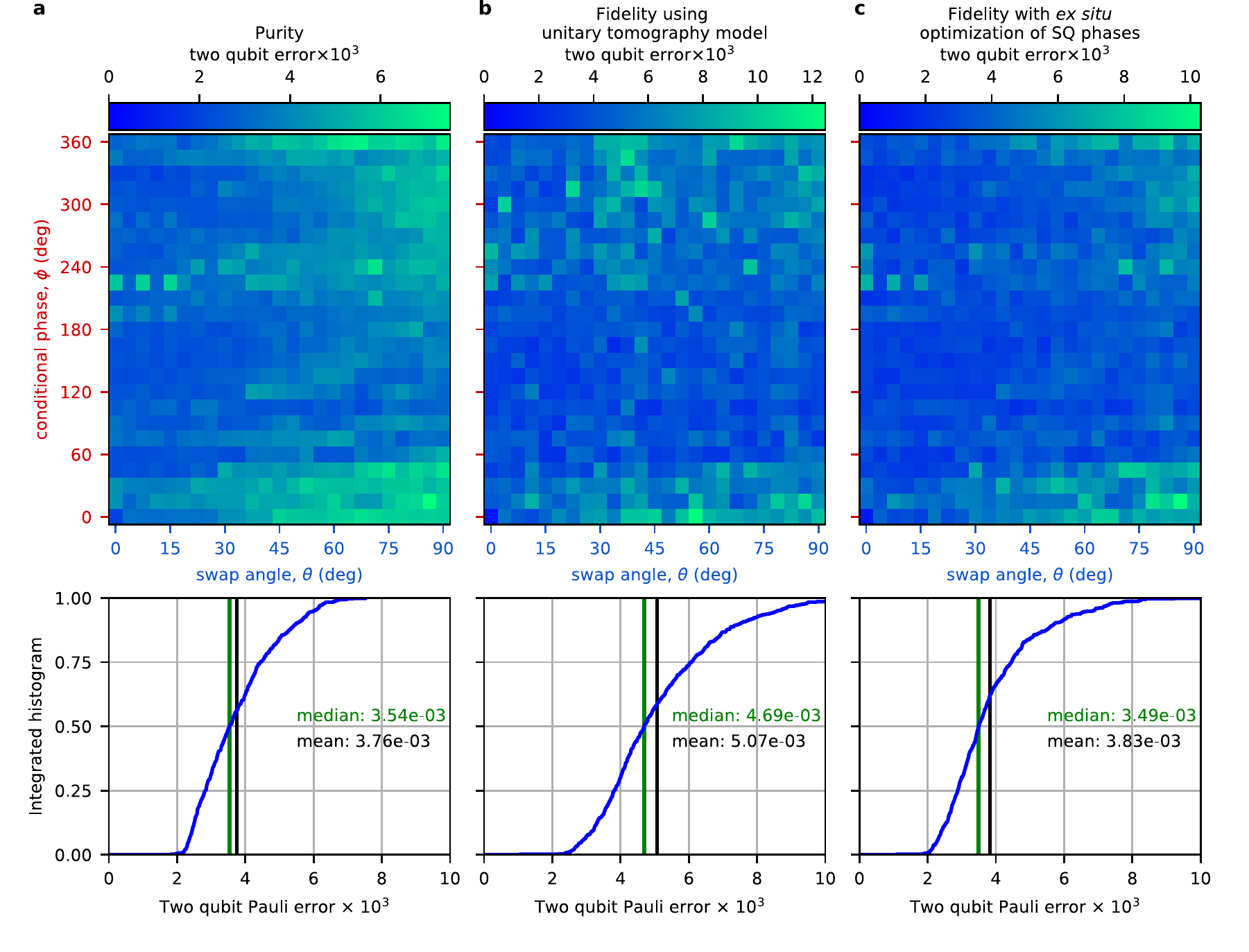}
\caption{\label{fig:purity_exsitu}Comparison of purity benchmarking and cross-entropy benchmarking with and without a constrained \textit{ex situ} optimization of the fSim control angles. \textbf{a}, Purity benchmarking. \textbf{b}, XEB error per gate using the fSim gate model obtained from unitary tomography (supplement \ref{app:unitary_tomography}). \textbf{c}. XEB error after a constrained \textit{ex situ} optimization of the fSim gate parameters where $\theta$ and $\phi$ were held fixed to the grid and the single-qubit phases were optimized.  }
\end{figure*}

In Figure \ref{fig:purity_exsitu}a we perform Purity benchmarking for each XEB gate sequence and obtain an average Purity of $3.76 \times 10^{-3}$ per fSim gate. In Figure \ref{fig:purity_exsitu}b we plot $e_{p,\rm{2q,unitary\_tomography}}$, the Pauli error per fSim gate using the fSim gate model obtained from unitary tomography. The average $e_{p,\rm{2q,unitary\_tomography}}$ is $5.07 \times 10^{-3}$ indicating a coherent error of $1.31 \times 10^{-3}$ per fSim. In Figure \ref{fig:purity_exsitu}c we perform \textit{ex\,situ} optimization of our fSim gate model to reduce the coherent error by changing the three single-qubit detuning model parameters. We hold the values of $\theta$ and $\phi$ fixed to the sampling grid, but allow the single-qubit phases in the fSim model to be optimized. With this improved gate model coherent error is nearly eliminated. The average error $e_{p,\rm{2q,ex\,situ}}$ is $3.83 \times 10^{-3}$ reducing the average coherent error to $7 \times 10^{-5}$ per gate.

\begin{figure}
\includegraphics[width=89mm]{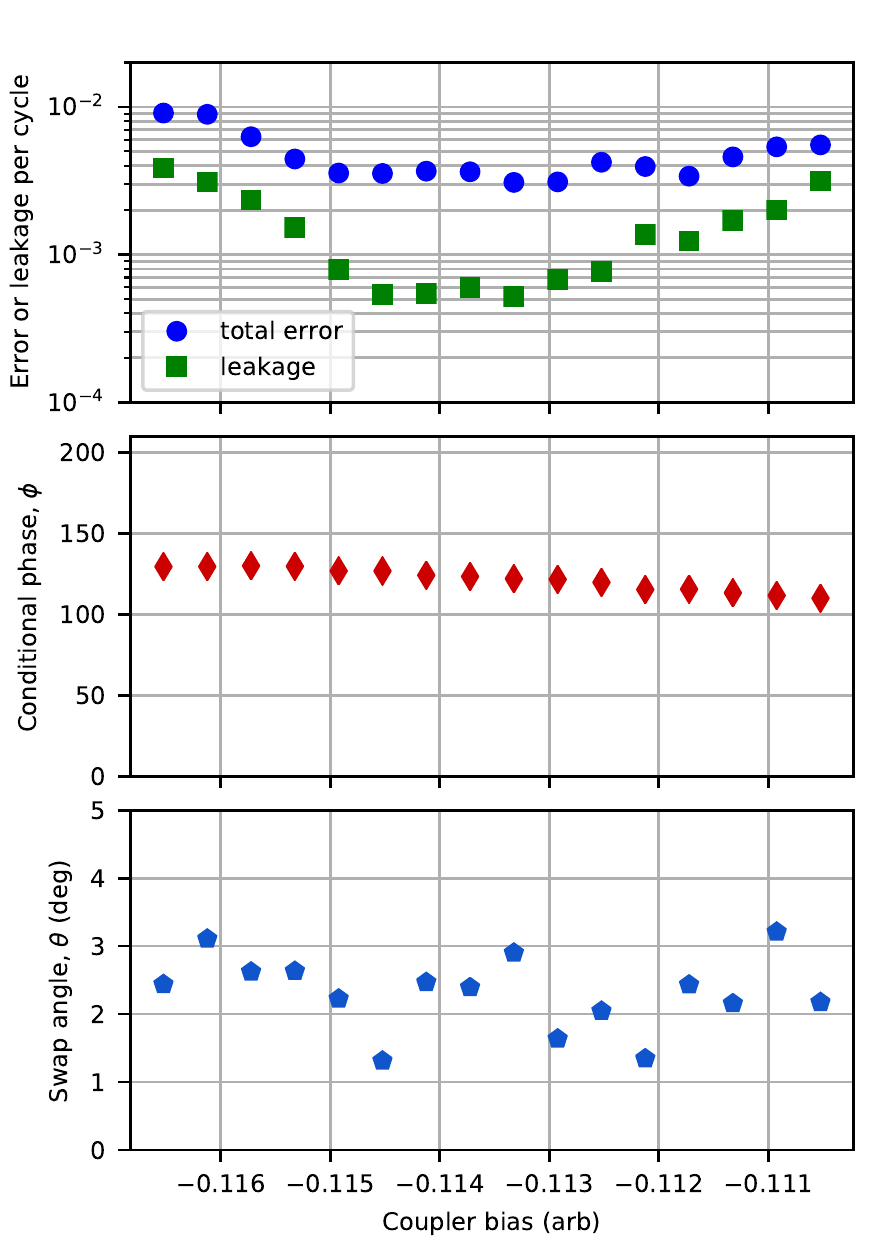}
\caption{\label{fig:leakage}Plot of leakage and total XEB error per cycle for a 13\,ns CPHASE gate as a function of the coupler bias. The increment on the x-axis is twice the minimum increment of our DAC ($2/2^{14}$, e.g. a 14-bit bipolar DAC $\approx 0.0002$). In this case we find that leakage reaches a minimum of $5-6\times 10^{-4}$ for a range of coupler amplitudes approximately $10x$ our minimum DAC adjustment.}
\end{figure}

We characterize leakage by directly measuring the $\ket{2}$ state population as a function of the XEB sequence depth. In Figure \ref{fig:leakage} we perform this measurement for a line cut of fSim control pulses that sweep the coupler bias on either side of the low-leakage bias used to perform a CPHASE gate. We find leakage to be minimized to a value of $5-6 \times 10^{-4}$ for a range of coupler biases spanning nearly 10 ``clicks'' of our 13-bit bipolar DAC ($2/2^{13} \approx 0.0002$).

In total, these metrics indicate that we have achieved incoherent-error-limited gates with fairly low leakage (if necessary, leakage may be reduced further by optimizing the gate length). Additionally, we find that we are able to perform the desired $\rm fSim(\theta, \phi)$ gate we want without incurring additional coherent error. A critical component in achieving these results was eliminating the non-gate-like behaviors induced by long settling tails on our flux bias pulses. As such, we will now detail the procedure used to calibrate our flux control pulses.

\subsection{\label{app:unitary_overlap}Unitary overlap}
\begin{figure}
\includegraphics[width=85mm]{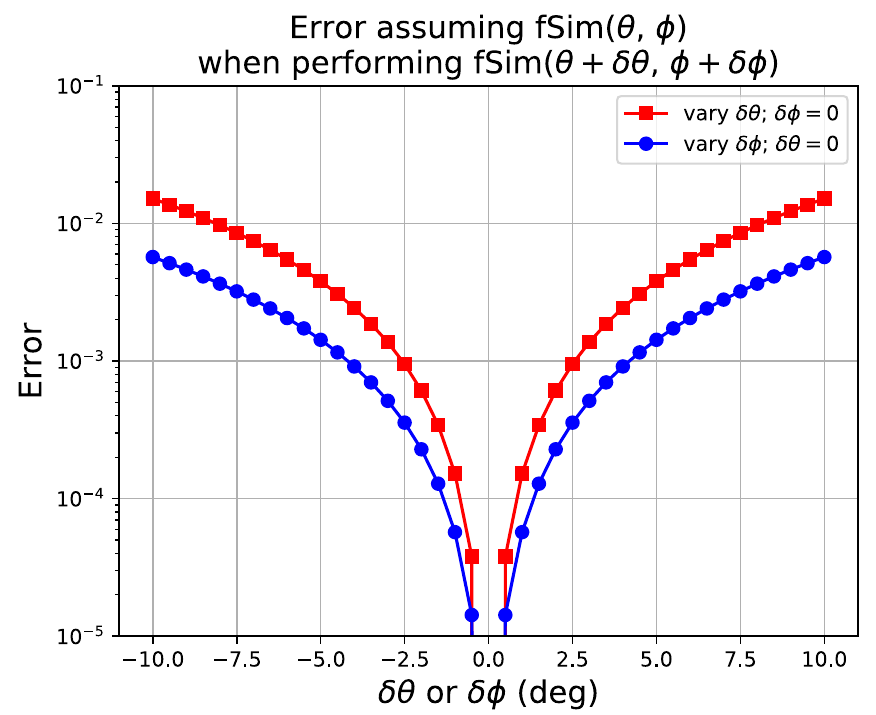}
\caption{\label{fig:unitary_overlap}We may choose to interpret some $\rm fSim_{actual}  = fSim(\theta + \delta \theta, \phi + \delta \phi)$ as some $\rm fSim_{target} = fSim(\theta, \phi)$, by accepting additional coherent error. For small deviations in either $\theta$ or $\phi$ the error is proportional to the square of the deviation.}
\end{figure}

The unitary overlap of two unitary matrices, e.g. some target fSim, $U_{\rm{target}}$, and the actual fSim, $U_{\rm{actual}}$, is defined as $Tr(U_{\rm{target}}\cdot U_{\rm{actual}})/D$, where $D$ is the dimension of the Hilbert space. The unitary overlap is related to the Pauli error, $e_p = 1-(Tr(U_{\rm{target}}\cdot U_{\rm{actual}})/D)^2$. The Pauli error in an fSim gate for small deviations in either $\theta$ or $\phi$ is proportional to the square of the deviation angle. In Figure \ref{fig:unitary_overlap} we plot the additional coherent error incurred if you assume some actual $\rm fSim_{actual} = fSim(\theta + \delta \theta, \phi + \delta \phi)$ is instead some target $\rm fSim_{target} = fSim(\theta, \phi)$. This plot indicated that a deviation of either $2.5^{\circ}$ in $\theta$ or $4^{\circ}$ in $\phi$ with result in an additional coherent error of $1 \times 10^{-3}$. In our case (Figure \ref{fig:purity_exsitu}), after a constrained optimization where $\theta$ and $\phi$ were fixed to a grid, our average error was approximately $1 \times 10^{-4}$ higher than the purity limit which corresponds to a deviation of about $1^{\circ}$ in either $\theta$ or $\phi$.

\section{\label{supplement:pulse_calibration}Control pulse calibration}
In a world without flux settling tails, we would be able to implement an arbitrary fSim gate with a fidelity that is the sum of the requisite CPHASE and iSWAP-like gates by just merging the control pulses into a composite fSim gate. Unfortunately, due to flux settling tails, further calibration, described in \ref{app:fsim_calibration}, was required. The keystones of this calibration were two-fold: 1) When performing two flux control based gates back to back (e.g. 2\,ns separation), adjust the amplitude of the second pulse based on the first. 2)When implementing a composite gate, perform a CPHASE gate followed by the iSWAP-like gate so that bleed through is well behaved; in the reverse order, bleed through of the iSWAP-like coupler pulse into the CPHASE gate pulses will result in leakage to the $\ket{2}$ state which is an error in the fSim model. Using these two principles, we were able to implement a robust calibration of the complete fSim gate set.

As we have demonstrated numerically in supplement \ref{app:numerics}, our desired implementation of the fSim gate set is possible with less than 1\% error using simple rectangular control pulses. Unfortunately, the system transfer function (electronics and wiring) is imperfect and cannot produce these ideal waveforms exactly. Fortunately, as explored numerically in Figure \ref{fig:sim_smooth}, our fSim implementation is mostly sensitive to the integral of our control pulses rather than the shape. This likely remains true unless the spectral content of our flux control pulses approaches the qubit frequency. However, we must be very careful to ensure our control pulses do not bleed into each other which requires careful calibration of our flux bias settling tails.

We can consider settling non-idealities at two time scales: 1) pulse distortion during the duration of a gate (roughly 15\,ns), and 2) pulse settling that occurs after the intended gate duration. Distortion at short times may, for instance, make it difficult to place the qubits exactly on resonance during a gate\textemdash this may make it difficult to achieve a swap angle, $\theta$, of $90^{\circ}$ swap amplitude (Rabi oscillation amplitude $= g^2/(g^2 + \pi \Delta^2/2) = 1$ if and only if the qubits are on resonance), but fortunately these distortions do not have a huge impact on the rest of the fSim parameter space. Due to the periodic nature of Rabi oscillations the resulting fSim is mostly dependent on the integral of the control pulses. Pulse settling that occurs outside the intended gate interval means that adjacent gates will bleed in to each other. If the tails are relatively short (a few ns), it is possible to mitigate this error just by placing a short idle time between gates. Pulse settling at longer times is particularly nefarious because it becomes no longer feasible to pad gates with idle times and setting times of 5-1000+\,ns have been observed in superconducting qubit systems. If left uncompensated, the performance of the $m^{th}$ 15\,ns long gate would be dependent on the preceding 1-60+ gates. This runs contrary to the entire notion of gate-based local operations and certainly would not fit within our static fSim control model used with XEB. As such, it is this long-time settling in particular that requires a careful calibration to enable the sensible control strategy employed throughout this letter. 

The full fSim gate calibration happens in three stages.  In the first stage, we calibrate the electronics to eliminate the long-time settling flux settling. In the second stage, we describe the calibration procedure for the CPHASE and iSWAP-like gate sets. Then, for the fSim gate family, we perform further calibrations of the composite fSim gates to achieve the best possible gate performance by adjusting the control amplitude of the second pulse dependent on the first rather than adding longer buffer times between flux pulses.

\subsection{\label{app:electronics_calibration}Electronics calibration}

On this device there are a total of seven flux bias lines, four for the qubits and three for the couplers.  Each channel is driven by a dedicated 1\,GS/s, 14-bit DAC controlled by an FPGA to form an arbitrary waveform generator.  Each line uses nominally identical cabling, attenuation, and filtering from room temperature down to the sample's chip mount. To compensate for non-idealities in each line, we first measure the qubit's response to a flux pulse, fit the response using three exponential decay time constants, and then use this model to pre-distort our control pulses as in previous work \cite{Barends2014, foxen2018}. This allows us to directly measure and compensate for the transfer function of each qubit's flux bias wiring. Implementing a similar \textit{in situ} calibration of the coupler bias lines is the subject of on-going work. For now, we have found it sufficient to simply apply the average of the two adjacent qubit settling models to the coupler. The pulse calibration parameters for the pair of qubits and the coupler used to benchmark the fSim gate set are summarized in Table \ref{table:settling_params}.

\begin{table}
\caption{\label{table:settling_params}Summary of the settling parameters for two qubits.  The average of the settling compensation for these two qubits was applied to the coupler.}
\scriptsize

\begin{tabular}{|c|c|c|c|c|c|c|}
\hline
& $\alpha_1$ (\%) & $\tau_1$ (ns) &  $\alpha_2$ (\%) & $\tau_2$ (ns) & $\alpha_3$ (\%) & $\tau_3$ (ns)\\ \hline
$q_2$ & -0.46 & 858 & -1.00 & 104 & -4.94 & 10 \\ \hline
$q_3$ & -0.61 & 996 & -0.82 & 94 & -5.97 & 9 \\ \hline
coupler (avg $q_2$ \& $q_3$) & -0.53 & 927 & -0.91 & 99 & -5.45 & 10 \\ \hline
\end{tabular}\\

\end{table}
\normalsize

After performing the electronics calibration we find the unitary gate interactions of our fSim gates to be well characterized by either unitary tomography, performed with a depth-1 circuit, or cross-entropy bench-marking using a depth N circuit where N varies from 5 to 700.  This fact is illustrated by Figure \ref{fig:purity_exsitu} panels b and c where the difference in the average error of all 525 fSim gates differs by only $\rm 1.2 \times 10^{-3}$ with and without optimizing the single-qubit unitary parameters\textemdash this provides an upper bound on the effects of pulse bleed through on gate fidelity. If we consider the gate timing of the cross-entropy benchmarking sequence in Figure \ref{fig:purity_exsitu}, which used 28\,ns fSim gates interleaved with 15\,ns single-qubit gates, this result indicates that our settling is well compensated for at times longer than 15\,ns. This result also indicates that the qubit biases are settled enough to have a minimal impact on the single-qubit gate errors.  If this were not the case then we would require a circuit-depth-dependent gate model to reach the purity limit. However, the settling of the coupler bias flux signal at times less than 15\,ns becomes non-negligible and merits special consideration when calibrating fSim gates composed of a CPHASE gate in close proximity to an iSWAP-like gate. So, we will first detail the calibration procedure for each of the component gate families in the next section and finish our calibration discussion with a description of composite fSim calibration procedure. 

\subsection{\label{app:cphase_iswap-like}CPHASE and iSWAP-like calibrations}
\begin{figure}
\includegraphics[width=89mm]{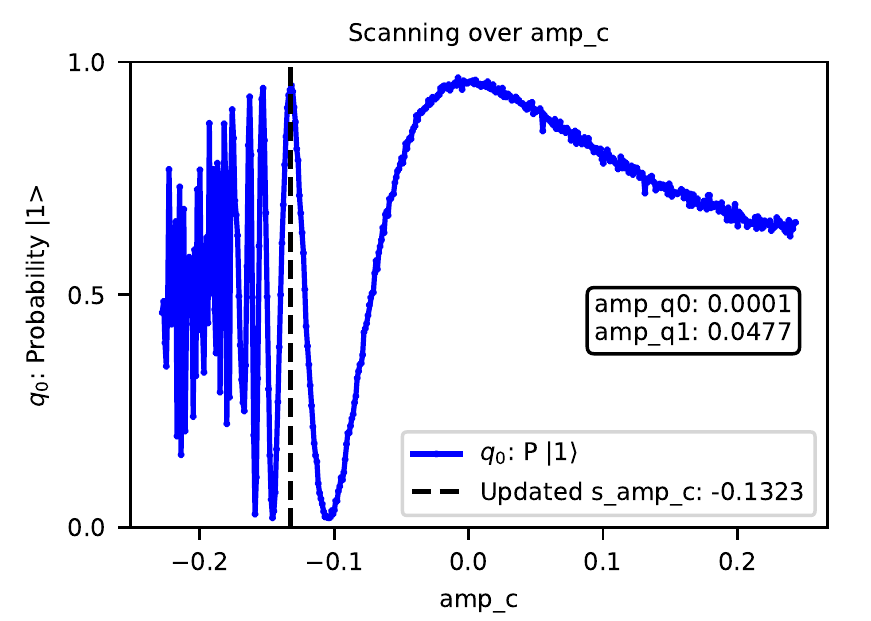}
\caption{\label{fig:cphase_cal} CPHASE gate calibration scan. We use the pre-calibrated bias-to-qubit frequency function to choose a desired qubit-qubit detuning and then sweep the amplitude of the coupler bias, $amp_c$ to identify the amplitude that completes a diabatic $\ket{11} \leftrightarrow \ket{02}$ swap indicated by the dotted vertical line.}
\end{figure}

We calibrate the CPHASE interaction by repeating the leakage experiment described in Figure \ref{fig:background_scans}a to fine tune the coupler bias amplitudes and to identify combinations of qubit detunings, $\Delta$, and corresponding coupler bias amplitudes that yield low-leakage gates. We use the qubit frequency bias transfer function to choose qubit biases that set the desired qubit detuning, $\Delta$ in the vicinity of $\eta$.  The frequency range around $\eta$ is set by the width of the Rabi interaction which, for a fixed pulse length, is inversely proportional to gate length since shorter gates require stronger coupling, $g$ (see Figure \ref{fig:sim_length} in supplement \ref{app:numerics}). We use 15\,ns pulses (13\,ns rectangular pulses with a 1\,ns padding on either side) which makes the Rabi interaction span about 75\,MHz on either side of the qubit nonlinearity, $\eta$. For each detuning in this range, we repeat the experiment in Figure \ref{fig:background_scans}a varying the coupling strength to minimize leakage. An example of the raw data from this experiment is provided in Figure \ref{fig:cphase_cal} where the dotted line indicates the low-leakage coupler bias amplitude that achieves one full swap from $\ket{11}$ to $\ket{02}$ and back. We initialize the $\ket{11}$ state, apply the CPHASE control pulses, and measure the $\ket{1}$ state population of the lower frequency qubit to identify when the population has completed a full swap. Then, for each combination of $\Delta$ and the corresponding low-leakage coupler bias we repeat the experiment from Figure \ref{fig:background_scans}b to measure the conditional phase. This procedure works well for $\rm \phi:[-130^{\circ},130^{\circ}]$ until the Rabi swap amplitude becomes small and the peak is broad along the coupling strength line-cut. At that point, we extrapolate towards the zero coupling bias while measuring the conditional phase to fill out the rest of the conditional phase control space.

\begin{figure}
\includegraphics[width=87mm]{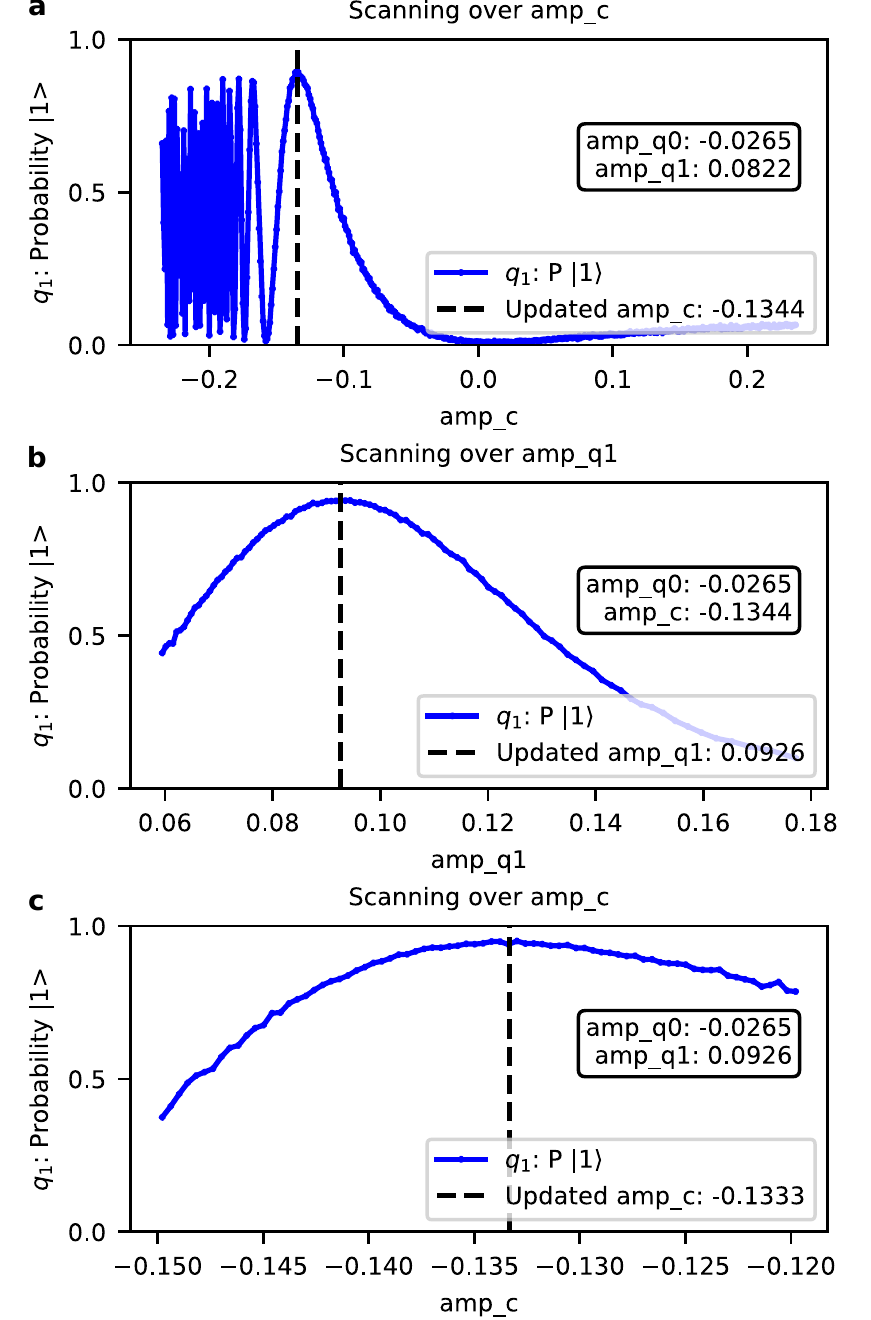}
\caption{\label{fig:iswap_like_cal} iSWAP-like gate calibration for $\theta = 90^{\circ}$ performed in three steps. In each experiment we initialize the $\ket{01}$ state and measure the population of the first qubit to identify the control biases to complete a full swap to $\ket{10}$ \textbf{a}, We use prior calibrations to bias the qubits on-resonance and scan the coupler bias amplitude, $amp\_c$, to find the bias that completes one swap. \textbf{b}, Using the new $amp\_c$, we scan the bias of one qubit, $amp\_q$, to place them on resonance. \textbf{c}, Using the updated qubit biases, we again scan $amp\_c$ to find tune the coupler bias.}
\end{figure}

In Figure \ref{fig:iswap_like_cal} we calibrate a 13\,ns iSWAP-like gate (11\,ns rectangular pulses, with 1\,ns padding) by repeating the experiment from Figure \ref{fig:background_scans}c three times with the qubits on resonance ($\rm \Delta = 0\,$MHz) to fine-tune the pulse amplitudes needed to reach $\rm \theta = 90^{\circ}$. Then, for $0^{\circ} < \theta < 90^{\circ}$ we simply interpolate the coupler bias between the ``OFF'' bias and the $\rm \theta = 90^{\circ}$ bias. For each iSWAP-like tune up experiment we initialize one qubit to the $\ket{1}$ state, apply the iSWAP-like pulses to the qubits and coupler, and then measure the $\ket{1}$ state population of the other qubit. In Figure \ref{fig:iswap_like_cal}a, we first use our pre-calibrated qubit frequency bias DC transfer functions to choose qubit bias amplitudes, $amp\_q0$ and $amp\_q1$, that place both qubits at the same frequency, and we sweep the coupler bias from the ``OFF'' bias to the maximum coupling bias to identify the amplitude that achieves exactly one a swap from the first to the second qubit corresponding to $\rm \theta = 90^{\circ}$ (dotted line). In Figure \ref{fig:iswap_like_cal}b, we repeat the experiment using the $\rm \theta = 90^{\circ}$ coupler bias from \ref{fig:iswap_like_cal}a while sweeping the bias of one qubit to maximize the amplitude of the swapped population, thus placing the qubits on resonance. Finally, in \ref{fig:iswap_like_cal}c, we repeat the experiment using the new qubit biases to fine-tune the coupler bias.

\subsection{\label{app:fsim_calibration}fSim calibration}
\begin{figure*}
\includegraphics[width=160mm]{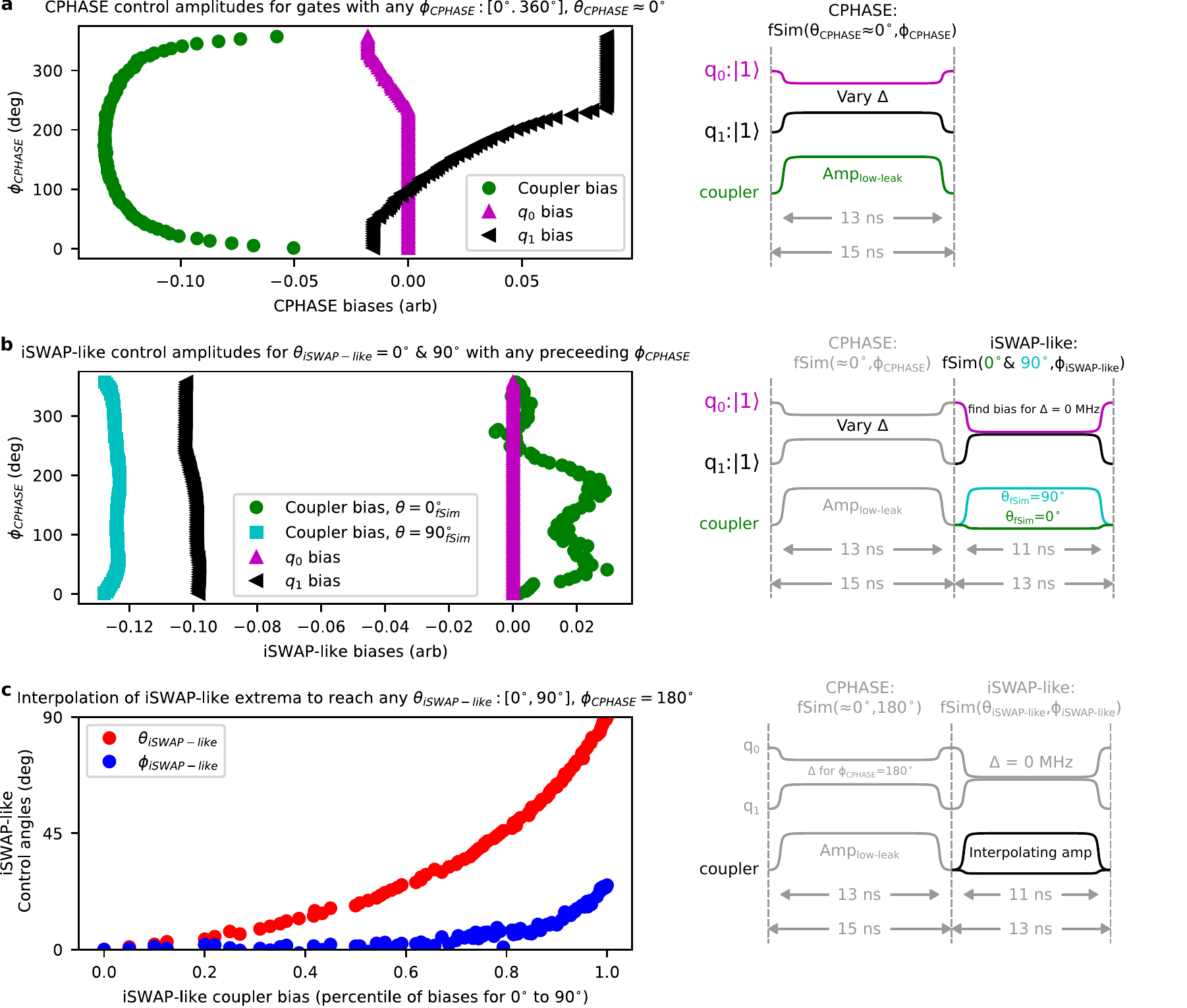}
\caption{\label{fig:fsim_cal_steps} Three steps to calibrating the fSim gate to account for pulse distortion of the first (CPHASE) pulses bleeding into the second (iSWAP-like) pulses. \textbf{a}, Follow the usual CPHASE calibration procedure to bring-up a full CPHASE gate family corresponding to fSim($\theta \approx 0^{\circ}, \phi:[0^{\circ}, 360^{\circ}]$). \textbf{b}, Follow the iSWAP-like tune up procedure, but play the CPHASE control pulses before the iSWAP-like pulses. Use the sequence to identify the flux bias amplitudes that achieve fSim gates with $\theta = 0^{\circ}$ and $\theta = 90^{\circ}$ for each proceeding CPHASE gate. \textbf{c}, for a preceding CPHASE gate with $\phi = 180^{\circ}$, bring up gates corresponding to fSim($\theta:[0^{\circ}, 90^{\circ}], \phi = 180^{\circ}$).}
\end{figure*}

Once we have calibrated the iSWAP-like and CPHASE gates, we should nominally be able to use one of each to implement any fSim gate.  Unfortunately our pulse response is imperfect at short times, as described in supplement \ref{app:electronics_calibration}. This was less of an issue for the iSWAP-like and CPHASE gates in section \ref{sec:cphase_iswap_like} because the XEB gate sequence alternated (10\,ns) single and (13\,ns or 15\,ns) two-qubit gates; in those cases, the uncompensated flux bias settling tails resulted in a small detuning at the qubit idle frequencies. However, when we perform an fSim gate as a composition of a CPHASE followed immediately by an iSWAP-like gate, the tail of the first coupler pulse bleeds into the second coupler pulse. Even a small settling tail adding to the amplitude of the coupler pulse can drastically change the coupling during the second gate due to the large coupler flux sensitivity at strong couplings (remembering Figure \ref{fig:coupling}c). In the future, this problem may be mitigated by identifying and removing the physical origin of these settling tails, with a more thorough \textit{in\,situ} calibration procedure for the couplers, or by placing longer idle times between gates.

In this work, we deal with pulse bleed through by calibrating composite fSim gates where the amplitudes of the second set of pulses in the composite fSim sequence is dependent on the first, thus eliminating the need for excessive idle times between gates. Conveniently, the tune up procedure for each gate in the composition is the same as in the isolated iSWAP-like or CPHASE case, just with the two sets of pulses played back-to-back\textemdash this works because each experiment in our usual bring-up procedure operates within an isolated manifold (e.g. one excitation for $\rm \theta$ or two excitations for $\rm \phi$) when performing fSim gates. The ordering of the gates within the fSim gate is chosen to place the CPHASE gate before the iSWAP-like gate. Since both coupler pulses have the same sign, if the CPHASE coupler amplitude bleeds into the iSWAP-like coupler amplitude, this results in slightly more swapping which is easily measured and adjusted for by reducing the iSWAP-like coupler amplitude to compensate. If we ordered the gates in the reverse order, pulse bleed through would generate leakage during the CPHASE gate which is much more difficult to characterize and remove.

For the purpose of building a robust registry of gates, we erred on the side of over-calibration for this demonstration.  However, we find these control parameters to be well behaved and it should be possible to sample more sparsely in the future to simplify calibration of the full fSim gate set. Figure \ref{fig:fsim_cal_steps} outlines the three steps used to calibrate our composite fSim gates. In Figure \ref{fig:fsim_cal_steps}a, we first calibrate many CPHASE gates spaced every $\rm 1^{\circ}$ using control pulses for just the CPHASE gate as shown on the right following the procedure outlined in supplement \ref{app:cphase_iswap-like}. Then, in Figure \ref{fig:fsim_cal_steps}b, for each preceding CPHASE gate we follow the iSWAP-like calibration procedure (also supplement \ref{app:cphase_iswap-like}) to identify qubit and coupler bias amplitudes to achieve both a $\rm \theta = 0^{\circ}$ and $\rm 90^{\circ}$ gate. Finally, in Figure \ref{fig:fsim_cal_steps}c, for a CPHASE conditional phase, $\rm \phi_{CPHASE} = 180^{\circ}$ we tune up iSWAP-like gates for $\rm \theta$ from $0^{\circ}$ to $90^{\circ}$ in $1^{\circ}$ increments by interpolating between the min and maximum amplitudes determined in \ref{fig:fsim_cal_steps}b. We use this calibration to produce a spline for $\theta_{\rm{iSWAP-like}} \rightarrow{} \%(bias_{90^{\circ}}-bias_{0^{\circ}}$) and another for $\theta_{\rm{iSWAP-like}} \rightarrow{} \phi_{\rm{iSWAP-like}}$. 

With the fSim gate registry in hand we set out to benchmark specific fSim gates. For a given target fSim, we first look up the iSWAP-like pulse amplitudes that achieve the correct swap angle $\theta_{\rm{iSWAP-like}}$, and subtract the conditional phase due to the iSWAP-like gate from the total target to choose pulse amplitudes for the desired CPHASE gate (e.g. $\phi_{\rm{CPHASE}} = \phi_{\rm{target}} - \phi_{\rm{iSWAP-like}}$). We then performed unitary tomography (supplement \ref{app:unitary_tomography}) using the pulse amplitudes we looked up in the registry to quickly assess the resulting fSim control angles of the composite gate. If either control angle is off by more than $1^{\circ}$, we used the registry to adjust the corresponding iSWAP-like ($\rm \theta$) or CPHASE ($\rm \phi$) control amplitudes by $\pm 1^{\circ}$ accordingly. This process converged to an fSim gate within $1^{\circ}$ of both $\theta_{target}$ and $\phi_{target}$ for the target fSim with fewer than 9 adjustments for each of the 525 fSim gates we benchmarked. Once the unitary tomography experiment indicated the composite fSim gate produced a unitary operation near the target fSim gate, we performed purity and cross-entropy benchmarking.

\section{\label{app:stability}System stability}
As the size of quantum processors grows (number of qubits), so too does the time it takes to calibrate a device (at least until fully parallel calibrations are possible). As the system drifts from these calibrations over time, the performance of a processor will fall and calibrations must be revisited. If the required calibration time is long compared to the scale of drift, then the device becomes unusable in practice. While electronics drift with both time and temperature must be considered when designing a system, one particularly worrisome issue is the time dependence of two level system (TLS) defects entering and leaving the qubit spectrum \cite{Klimov2018}. 

Here we present a promising snapshot of the stability of our system. In the process of calibrating the fSim gate set, we started by calibrating the single-qubit gates and readout. We then operated with the same single-qubit calibrations for several days while we were working on the fSim gate ultimately obtaining our primary fSim benchmarking dataset about a week after the initial single-qubit calibration.  Shortly after this, a TLS showed up near one of the qubit's idle frequencies significantly limiting its coherence. Then, after about another week, we returned to the original calibration parameters to benchmark a subset of the same $\rm fSim(\theta, \phi)$ gates presented in the main text. We were pleasantly surprised to find that the original calibration was still good enough to produce high-fidelity gates.

\begin{figure*}
\includegraphics[width=170mm]{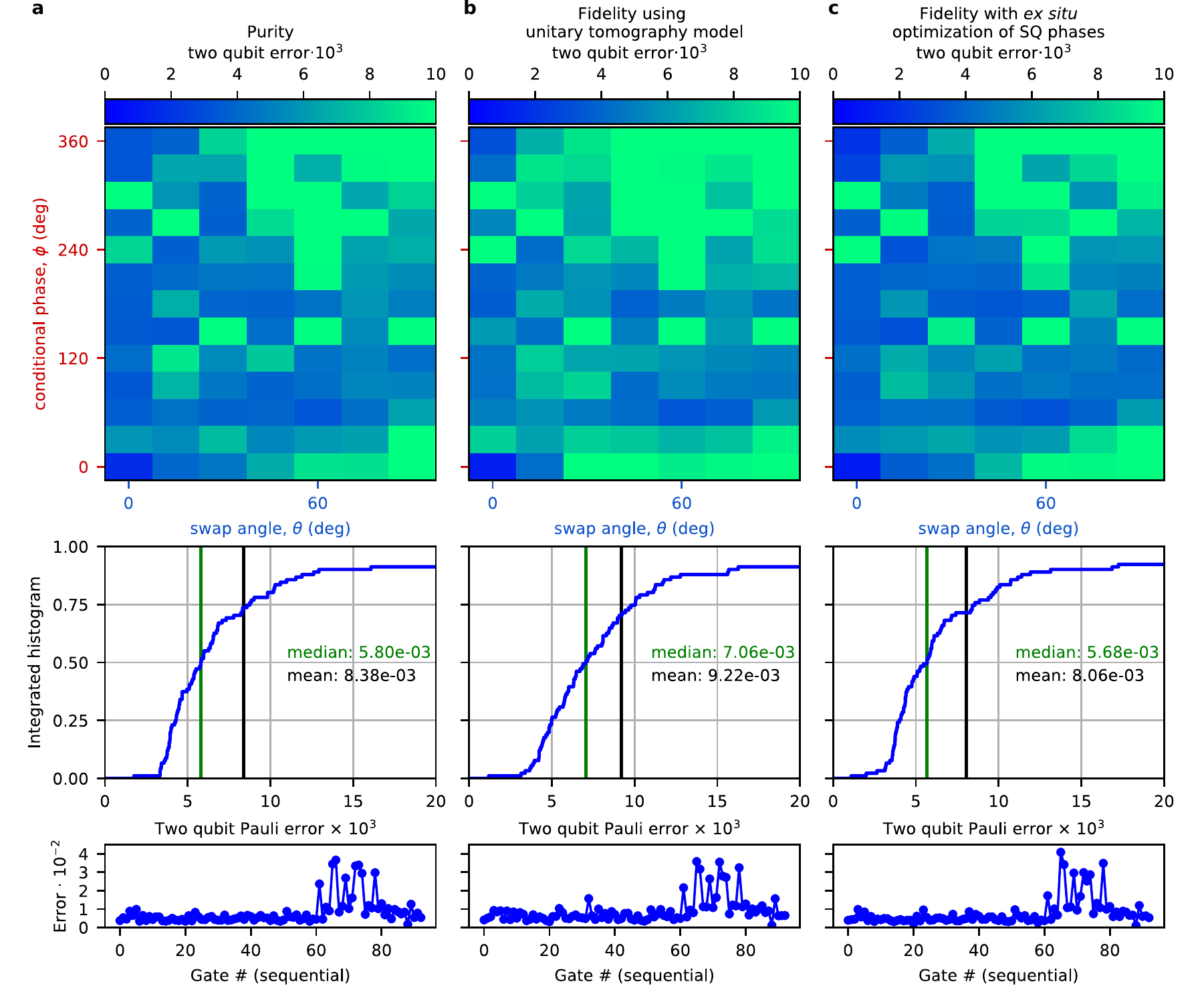}
\caption{\label{fig:stability}Snapshot of system stability. A few days after taking the data in Figure \ref{fig:fidelity}, a TLS showed up at one of the qubit idle frequencies effectively breaking the calibration. After about another week, we returned to the original calibration and repeated the fidelity measurement on a subset of 91 fSim gates which we present here (\textbf{top} and \textbf{middle} rows). We find that the average gate fidelity had decreased somewhat, but is still above 99\%. Furthermore, if we look at the gate error rates sorted in the order they were measured (\textbf{bottom} row), a strong time-dependence becomes apparent. Many of the gates presenting low errors ($\approx 5 \times 10^{-3}$) as they did after the initial calibration. It is not until gates numbered 60 to 80 or so where large errors show up. This indicates that our control electronics are stable enough to maintain a high-fidelity calibration on the timescale of weeks, and that TLSs are likely the biggest threat to maintaining long term calibrations.}
\end{figure*}

In Figure \ref{fig:stability} we used the two-week-old iSWAP-like and CPHASE calibrations to benchmark a less dense grid of fSim gates.  While the average performance has degraded by a factor of two from the initial calibration, the average error is still less than 1\%, but that is not the whole story. These fSim gates were benchmarked in a random order\textemdash if we look at a plot of the gate error as a function of time for these 91 (figure \ref{fig:stability}b), we see a strong time dependence where the first 50 gates (benchmarked over the course of an hour) have an average error much lower than gates \#50 to \#80.  This would seem to indicated that the two-week-old electronics calibration is stable enough to maintain high-fidelity gates for weeks, and that the decreased fidelity is likely due to the residual and/or intermittent presence of a TLS interacting with one of the qubits. In an ideal world, we would be able to prevent or remove TLS defects, but, at least presently, we do not know how to do this. Instead, relying on the stability of our electronics, an optimal strategy for maintaining up-time on a large-scale quantum processor will likely involve calibrating a number of idle frequency configurations and being able to quickly vet and switch to an old configuration if and when a TLS shows up.

\end{document}